# Wireless Sensor Network Virtualization: A Survey

Imran Khan, *Student Member, IEEE*, Fatna Belqasmi, *Member, IEEE,* Roch Glitho, *Senior Member, IEEE,* Noel Crespi, *Senior Member, IEEE,* Monique Morrow and Paul Polakos

*Abstract*— Wireless Sensor Networks (WSNs) are the key components of the emerging Internet-of-Things (IoT) paradigm. They are now ubiquitous and used in a plurality of application domains. WSNs are still domain specific and usually deployed to support a specific application. However, as WSNs' nodes are becoming more and more powerful, it is getting more and more pertinent to research how multiple applications could share a very same WSN infrastructure. Virtualization is a technology that can potentially enable this sharing. This paper is a survey on WSN virtualization. It provides a comprehensive review of the state-of-the-art and an in-depth discussion of the research issues. We introduce the basics of WSN virtualization and motivate its pertinence with carefully selected scenarios. Existing works are presented in detail and critically evaluated using a set of requirements derived from the scenarios. The pertinent research projects are also reviewed. Several research issues are also discussed with hints on how they could be tackled.

*Index Terms*— Wireless Sensor Network (WSN), Internet-of-Things (IoT), Virtualization, Node-level Virtualization, Network-level Virtualization

## I. INTRODUCTION

THE emerging Internet-of-Things (IoT) concept is considered to be the next technological revolution, one that realizes communication between many types of objects, machines and devices, and at an unprecedented scale [1]. WSNs can be seen as the basic constituents of IoT because they can help users (humans or machines) to interact with their environment and react to real-world events. These WSNs are composed of nodes that are amalgamations of micro-electro-mechanical systems, wireless communications and digital electronics, and have the ability to sense their environment, perform computations and communicate [2]. The most obvious drawback of the current WSNs is that they are domain-specific and task-oriented, tailored for particular applications with little or no possibility of reusing them for newer applications. This strategy is inefficient and leads to redundant deployments when new applications are contemplated. With the introduction of the IoT, it is not unrealistic to envision that future WSN deployments will have to support multiple applications simultaneously.

Virtualization is a well-established concept that allows the abstraction of actual physical computing resources into logical units, enabling their efficient usage by multiple independent users [3]. It is a promising technique that can allow the efficient utilization of WSN deployments, as multiple applications will be able to co-exist on the same virtualized WSN. Virtualization is a key technique for the realization of the Future Internet [4] and it is indeed quite pertinent to explore it in the context of WSNs.

Virtualizing WSNs brings with it many benefits; for example, even applications that were not envisioned a priori may be able to utilize existing WSN deployments. A second, related benefit is the elimination of tight coupling between WSN services/applications and WSN deployments. This allows experienced as well as novice application developers to develop innovative WSN applications without needing to know the technical details of the WSNs involved. Another benefit is that WSN applications and services can utilize as well as be utilized by third-party applications. It can also help to define a business model, with roles such as physical WSN provider, virtual WSN provider and WSN service provider.

The WSN virtualization concept can be applied to several interesting application areas. Recent advances in smart phones and autonomous vehicles [5] have made it possible to have multiple on-board sensors on them. Mobile crowd sensing is one area that can take advantage of virtualizing these sensors through participatory and opportunistic sensing [6] and [7]. An opportunistic urban sensing scenario is presented in [7] in which thousands of sensors are required to monitor the $CO_2$ concentration in an urban city. Instead of deploying these sensors and managing them, WSN virtualization can be used as a key enabling technology to utilize sensors from citizens to provide the required data. Similarly, Sensing-as-a-Service (SaaS) model is presented in [8] along with several use case scenarios. WSN virtualization can help realize the SaaS model through cost-efficient utilization of deployed sensors. Several

Manuscript received:

"This work is partially supported by CISCO systems through grant (CG-576719), European ITEA-2 funded project Web-of-Objects (WoO) and by the Canadian Natural Sciences and Engineering Research Council (NSERC) through the Canada Research Chair in End-User Service Engineering for Communications Networks.

I. Khan (imran@ieee.org) and N. Crespi (noel.crespi@it-sudparis.eu) are with Institut Mines-Télécom, Télécom SudParis, Evry, 91011, France.

F. Belqasmi (fatna.belqasmi@zu.ac.ae) is with Zayed University, Abu Dhabi UAE.

R. Glitho (glitho@ece.concordia.ca) is with Concordia Institute for Information Systems Engineering (CIISE), Concordia University, Montreal, H3G 2W1, Canada.

M. Morrow (mmorrow@cisco.com) and P. Polakos (ppolakos@cisco.com) are with CISCO Systems, Inc.

other motivational examples can be found in [9] and [10].

Of course there are many technical challenges to resolve before such utilization takes place but they also provide a strong motivation for a deeper and complete search space exploration to propose innovative solutions in this area. Many researcher now consider WSN virtualization as a key enabling technology and provide its motivation. According to the authors in [11], WSN virtualization is a powerful enabler for information sharing in the context of IoT by using it along with data analysis techniques. A smart city environment is considered in [12], where WSN virtualization could be used to efficiently utilize the deployed infrastructure. To achieve this type of utilization, the use of multiple concurrency models is advised, depending on the usage context. In [13], WSN virtualization is discussed as a key enabler to promote resource efficiency, with a cooperative model that captures several aspects of WSN virtualization. In [14] WSN virtualization is envisioned as an important technology to create large-scale sensor platforms that are used to satisfy efficient usage of network resources.

There are surveys (e.g. [15]) that cover wireless network virtualization at large, but they do not focus on the specifics of WSN virtualization. Although it is a key enabling technology, the few surveys published to date on WSN virtualization (e.g. reference [16], reference [17]), have several limitations. They do not include real world motivating scenarios and are also dated because they do not review the most recent developments in the area. Furthermore they lack comprehensiveness in terms of what is reviewed and how it is reviewed. There is for instance no well-defined yardstick for the critical analysis of the state of the art. In addition, they do not elaborate on potential solutions when it comes to research directions.

This paper is a survey on wireless sensor network virtualization. It aims at addressing the shortcomings of the very few surveys published so far on the topic. From that perspective it makes the following contributions:

- Real world motivating scenarios for WSN virtualization.
- Comprehensive and in-depth review of the state of the art including the most recent developments in the area.
- Critical analysis of the state of the art using well defined yard-sticks derived from the motivating scenarios.
- An overview of the open issues along with insights on how they might be solved.

In section II we discuss the basics of WSN virtualization concepts and its types. In section III, we first present the motivating scenarios and then provide a set of requirements. Based on these requirements we critically review the state-of-the-art in section IV. Relevant WSN virtualization projects are discussed in section V. Section VI outlines several research directions and section VII concludes the paper.

## II. WSN VIRTUALIZATION BASICS

WSN virtualization can be broadly classified into two categories: Node-level virtualization and Network-level virtualization. In this section we discuss both these categories.

### A. Node-level Virtualization

WSN node-level virtualization allows multiple applications to run their tasks concurrently on a single sensor node [18], so that a sensor node can essentially become a multi-purpose device. The basic concepts of node level virtualization are illustrated in figure 1. There are two ways to achieve node-level virtualization: Sequential and Simultaneous execution.

Sequential execution can be termed a weak form of virtualization, in which the actual execution of application tasks occurs one-by-one (in series). The advantage of this approach is its simple implementation, while the obvious disadvantage is that applications have to wait in a queue. In simultaneous execution, application tasks are executed in a time-sliced fashion by rapidly switching the context from one task to another. The advantage of this approach is that application tasks that take less time to execute will not be blocked by longer running application tasks, while the disadvantage is its complexity.

### B. Network-level Virtualization

It is WSN network-level virtualization that enables a Virtual Sensor Network (VSN). A VSN is formed by a subset of a WSN's nodes that is dedicated to one application at a given time [19]. Enabling the dynamic formation of such subsets ensures resource efficiency, because the remaining nodes are available for different multiple applications (even for applications that had not been envisaged when the WSN was deployed), although not necessarily simultaneously.

WSN network-level virtualization can be achieved in two different ways. One way is by creating multiple VSNs over the same underlying WSN infrastructure, as illustrated in Figure 2a. WSN nodes that are not part of any VSN remain available for other applications or network functions, such as routing. The second way is where a VSN is composed of WSN nodes from three administratively different WSNs, as shown in Figure 2b, facilitating data exchange between them that would not be possible otherwise.

## III. WSN VIRTUALIZATION – MOTIVATING SCENARIOS AND REQUIREMENTS

In this section we first present two scenarios that are derived from the literature, and then come up with a set of requirements. Using these requirements we critically review the existing work, grouping our summation of that work under three types: node-level virtualization, network-level virtualization and hybrid solutions.

### A. Motivating Scenarios

The scenarios described here illustrate the motivation and benefits of using WSN virtualization in common WSN deployments.

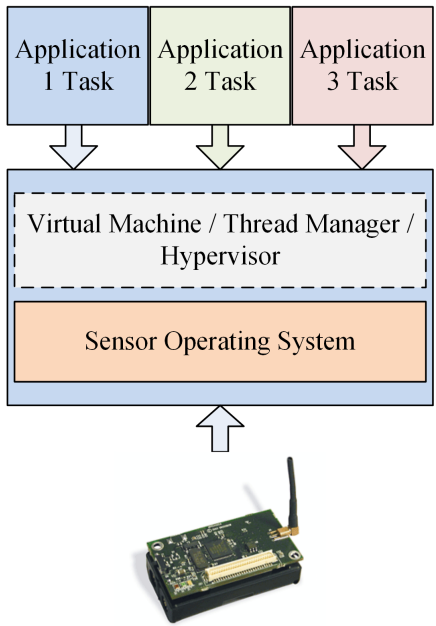

Fig 1: Execution of multiple applications in a general purpose WSN node

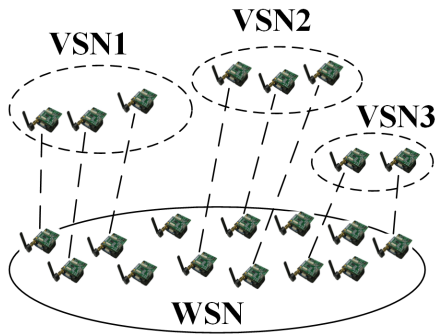

a) Multiple VSNs over single WSN

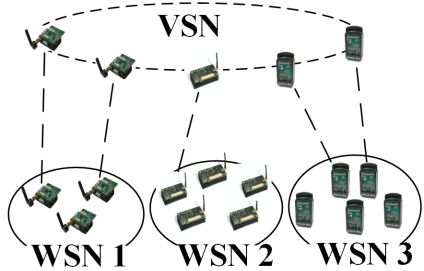

b) Single VSN over multiple WSNs

Fig 2: VSN concept

*1) Fire Monitoring Scenario*

Consider the example of a city near an area where brush fires are common [9]. We assume that the city administration is interested in the early detection of fire eruption and in its course, using a WSN and a fire contour algorithm to determine the curve, shape and direction of fire. One approach is that the city administration could deploy WSN nodes all over the city (i.e. on each street and at individual houses), but this is not very efficient because some individuals may have already deployed WSN nodes in their homes to detect fires. A more efficient approach would be for the city administration to deploy WSN nodes to areas under its jurisdiction, i.e. streets and parks, and to re-use the WSN nodes already deployed in private homes. In this scenario, two different applications share the same WSN infrastructure: one, belonging to home owners, is confined to private WSNs deployed in individual houses, and the other belongs to the city administration and shares the private WSN nodes with the WSN nodes deployed by the city administration. Periodic notification or query-based models are not suitable because the city administration application requires complete access to all the WSN nodes for adaptive sampling.

Another issue is that in order to execute a fire contour algorithm in a distributed fashion, WSN nodes need to exchange fire notification messages with each other. The query-based data exchange approach is not efficient as it will force the execution of the fire contour algorithm at a remote centralized location, since two WSN nodes located in their respective private domains cannot exchange data. An overlay network is one possible solution. This scenario illustrates the need for WSN virtualization, as two different users need to share a common resource, i.e. WSN nodes.

*2) Heritage Building Monitoring*

A real-world deployment of a WSN is presented in [20], in which a WSN is used to monitor the impact of constructing a road tunnel under an ancient tower in Italy, as it was feared that the tower could lose its ability to stand on its own and collapse during the construction. Now consider that there are three users interested in the fate of the tower. The first is the construction company, as it needs make sure that the tower does not lose its ability to stand on its own, otherwise it will have to pay a heavy fine. The second user is the conservation board that routinely monitors all the ancient sites around the city, and the third user is the local municipality which will have to plan emergency remedial/rescue actions in case the tower falls during the construction.

It is quite possible that the conservation board has already deployed its own WSN to monitor the health of ancient sites including this tower. In this case the construction company and the local municipality can use the existing sensor nodes during the construction period. In the absence of WSN virtualization, there are only two possible solutions. One is to rely on the information provided by the conservation board's application. However this information may not be at the required granularity level. Worse, some of the information that is needed might simply not be available because the requirements of the construction company and of the local municipality were not considered when the conservation board application was designed and implemented. The second solution is that each user deploys redundant WSN nodes.

*B. Requirements*

In this section we present a list of eight requirements, derived from the scenarios mentioned above. In Table IV we indicate if the existing solutions meet our identified requirements, and to what degree.

The *first* requirement is the availability of node-level virtualization. This is a fundamental requirement which ensures that the sensor nodes can support the concurrent execution of multiple applications.

The *second* requirement is network-level virtualization, which concerns the ability of sensor nodes to dynamically form groups to perform the isolated and transparent execution of application tasks in such a way that each group belongs to a different application.

The *third* requirement is support for application/service priority. It is our observation that most WSNs are deployed for mission-critical situations like security, fire monitoring, battlefield conditions and surveillance. In such situations, mission-critical applications/services should have prioritized execution mechanisms.

The *fourth* requirement is that any WSN virtualization solution should be platform-independent and thus should not depend on a particular hardware or software platform.

The *fifth* requirement is that the proposed solution should have a resource discovery mechanism, for both neighbor discovery and service discovery.

The *sixth* requirement is based on the applicability of the proposed solution to resource-constrained sensor nodes, including early generation sensor nodes. Mechanisms to allow legacy sensor nodes to become part of a WSN virtualization solution are also covered by this requirement.

The *seventh* requirement is heterogeneity, which means that the solution should be applicable to a variety of WSN platforms with different capabilities (e.g. processing power, memory). These platforms would include MICAZ, MICA2, Atmel AVR family, and MPS430 among others.

The *eight* requirement is the ability to select sensor nodes for application tasks. When multiple applications concurrently utilize a deployed WSN, selection of proper sensor nodes is very important because applications may have spatial and temporal requirements [21].

## IV. STATE-OF-THE-ART

In this section we present the state-of-the-art and analyze it critically. We categorize the existing work as Node-level virtualization, Network-level virtualization or Hybrid solutions. Hybrid solutions combine both node- and network-level virtualization. Each category is further classified based on the approaches used.

*A. Node-level Virtualization*

We group the Node-level virtualization approaches under two umbrellas: sensor operating system (OS) based solutions and Virtual Machine-/Middleware (VM/M) based solutions. In sensor OS-based solutions, the node-level virtualization is part of the sensor OS. In VM/M-based solutions, the node-level virtualization is performed by a component running on top of the sensor's OS.

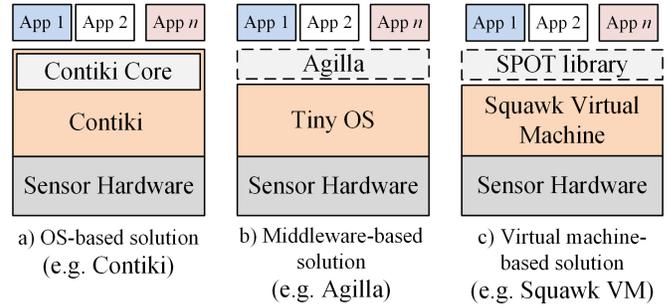

Fig 3: Example node-level virtualization solutions

Node-level virtualization solutions use two types of programming models; event-driven and thread-based. Event-driven programming model is simple to implement in sensors. Event-driven programs have a main loop that listens for the events, e.g. the temperature value going above a threshold. When the event occurs a callback function is called to handle the event, using an event-handler. When a program is blocked, by an I/O event, its event-handler simply returns the control without involving context switching. Thread-based model is more difficult to implement in sensors, due to limited resources and use of common address space. Each program consist of multiple threads, and when a thread is blocked, context switching is required to execute other threads [22].

Fig. 3 shows the node-level virtualization types while Table I illustrates the characteristics of the existing works addressing node-level virtualization.

*1) Sensor Operating System-based Solutions*

SenSmart [23] is a recent multitasking sensor OS that supports the execution of concurrent application tasks in very resource-constrained sensor nodes. It is designed to tackle the issues associated with the execution of concurrent application tasks. Normally, application tasks have their associated predefined stack space, but in SenSmart the stack allocation is managed dynamically at run time. Initially, each application task gets its default (stack) memory region and time slice, but during its execution SenSmart manages the size and location of the allocated stack in a transparent way. Each application task uses logical addresses at runtime, managed by the OS and mapped onto the physical memory. Stack space can be reclaimed from those tasks that no longer require it. When a new task is scheduled to run, the context of the current task is compressed and saved in a circular buffer for its resumption. The system architecture consists of a base station that compiles the code, links it and eventually distributes it to the sensor node. There is no mention of support for network layer support (6LoWPAN) or any radio protocol.

The support for node-level virtualization is provided by compiling and linking multiple application task codes together in a single code image. The application task codes are programmed in nesC and the compiled binary code of each task is then modified by a rewriter, combined with other binary codes and finally linked with the precompiled kernel runtime. The kernel runtime ensures that the application tasks, when instantiated, follow the multitasking semantics (stack management, context switching) and run concurrently. Once a final executable code is generated, it can be disseminated to the sensor node using any wireless reprogramming approach. The strategy of first compiling and linking all the binary codes together means that there is no separation of OS and application tasks, and, whenever a new application task is contemplated, all of the software of the sensor node is updated. The OS uses an event-driven programming model and follows a sense-and-send workflow model [24].

SenSmart has been implemented in Mica2/MicaZ hardware platforms and evaluated for overhead of common system functions, application benchmarking, and task scheduler performance when concurrent tasks are executed. The overhead of common system functions is within acceptable range especially for important functions such as context saving, restoring and switching. All these functions take between 127μs to 316μs. For application benchmarking it was found that the same applications use more CPU cycles in SenSmart than in TinyOS. For concurrent tasks, the evaluation found that delays recorded during execution of multiple tasks has same order of magnitude as context switching.

RIOT [25] is the latest attempt to address the challenges of designing a flexible OS for diverse hardware in the IoT. The concept of RIOT is based on the fact that none of the existing OSs, traditional or resource-constrained, are capable of supporting diverse hardware resources in the IoT. The focus of RIOT is to provide features such as real-time multithreading support, a developer-friendly programming model and POSIX-like API based on C/C++, as well as full TCP/IP network stack support for resource-constrained devices using 6LoWPAN and RPL. RIOT is based on microkernel architecture and requires only 1.5kB of RAM and 5kB of ROM for a basic application. RIOT can run on 8-bit, 16-bit and full 32-bit processors, and thus has the potential to become unique operating system for diverse hardware devices in the IoT paradigm. This adaptability is achieved by using a hardware abstraction layer. Overall, RIOT takes a modular approach and the system services and the user application tasks run as threads. The scheduler is designed to minimize context switching between threads to few clock cycles. The kernel is based on FireKernel [26] providing maximum reliability and real-time multithreading. System tasks have static memory allocation, but for application threads dynamic memory management is used. RIOT is a work in progress and so far there are no performance results or comparisons with existing OSs, but the code is available on their website.

In the context of WSN virtualization, RIOT uses a real-time thread-based programming model where various system services and application tasks are coded in standard ANSI C/C++ and run in parallel. Threads can be preempted based on their priority. Application tasks are coded independently of the hardware and software, which makes it possible to run them on different devices. In large-scale scenarios such as Smart Cities, sensor nodes and other IoT devices (e.g. surveillance cameras) can be programmed conveniently.

So far there are no performance results regarding RIOT OS however, in [27] the authors do present a theoretical comparison of their approach against existing competition without any qualitative or quantitative comparison.

SenSpire OS [28] is another recent effort that supports both event-driven and thread-based programming models. Their work has four main features: predictability – to guarantee that sensor nodes respond to control messages, availability – the nodes remain available for data forwarding when needed, programming mode – which is hybrid, and efficiency – so that the OS can be used on very resource-constrained sensor nodes. Another contribution of SenSpire is a multi-layer (*radio, resource and sensornet layers*) abstraction to develop networked applications. The radio layer makes it possible to write device drivers using different MAC protocols. The resource layer exposes the lower layer and allows different application tasks to use it concurrently. A new object-oriented language (CSpire) is provided to program user application tasks using a hybrid programming model. SenSpire uses static optimizations, meaning that application tasks, their states, and the kernel structures should be known beforehand. This limits its flexibility, a requirement for the real-world deployment of WSNs. The kernel of SenSpire is written in C and the application tasks are written in CSpire. The paper describes extensive results based on the implementation of SenSpire on Mica2, MicaZ, and TelosB nodes. Its performance at various benchmarks is compared to that of MANTIS [29] and TinyOS [30]. Overall findings indicate that SenSpire offers a performance comparable to those OSs.

For WSN virtualization, SenSpire incorporates both event-driven and thread-based programming models. Tasks can be programmed as events or as threads. Event tasks have higher priority than thread tasks. System tasks are usually implemented as event tasks because they are predictable and easier to maintain. Application tasks are implanted as thread tasks with varying priority levels. A thread task is preempted either by a higher-priority thread task or when it goes to sleep. This set up is unlike other OSs where thread tasks are executed in a time-sliced manner. In SenSpire the threads follow run-to-completion model unless they are preempted by a higher priority thread. The execution of threads is sequential (First-in First-out) when they have the same priority level. The use of CSpire language to program application tasks means a learning curve for developers. Despite using a layered-approach, application tasks are tightly integrated with the OS and so when new application tasks are contemplated, all of the sensor node software is updated.

The performance results of SenSpire OS show that its interrupt latency is less than TinyOS. The overhead of task scheduling is compared against MANTIS OS [29] showing more delay in case of SenSpire. The energy consumption of various tasks including radio and CPU are almost similar to TinyOS.

TABLE I
CHARACTERISTICS OF NODE-LEVEL VIRTUALIZATION SOLUTIONS

| Solution (Year) | Programming Model | Programming Language | Separation between OS and application tasks | Protocols Supported at different layers | Real-time Applications |
|---|---|---|---|---|---|
| SenSmart (2013) | Event-driven | nesC | No | Not discussed | No |
| RIOT (2013) | Thread-based | ANSI C/C++ | Yes | 6LoWPAN & RPL | Yes |
| PAVENET (2012) | Thread-based | C | No | Not discussed | Yes |
| SenSpire (2011) | Event- and thread-based | CSpire | No | CSMA, CSMA/CA, B-MAC & X-MAC | No |
| Nano-CF (2011) | Event-driven | Nano-CL | Yes | DSR, TDMA & B-MAC | Yes |
| UMADE (2010) | Event-driven | nesC | Yes | Not discussed | No |
| Agilla (2009) | Mobile agent and tuple-space -based | Assembly-like | Yes | Not discussed | No |
| LiteOS (2008) | Event- and thread-based | C | Yes | Not discussed | No |
| Squawk VM (2006) | Thread-based | J2ME | No | CTP, 6LoWPAN, AODV, LQRP | No |
| VMSTAR (2005) | Thread-based | Java | No | Not discussed | No |
| MANTIS (2005) | Thread-based | C | Yes | TDMA | No |
| TinyOS (2005) | Event-driven | nesC | No | Geographic routing, flooding, unicast | No |
| Contiki (2004) | Event- and thread-based | C | Yes | HTTP, COAP, UDP, TCP, RPL, 6LoWPAN | Only for event-driven applications |
| Maté (2002) | Event-driven | TinyScript | No | Not discussed | No |

MANTIS [29] is a thread-based embedded operating system supporting simultaneous execution on sensor nodes. The OS kernel and threads are programmed in C language and are portable across different hardware platforms. There are system-level threads and user-level threads. The OS kernel, scheduler and underlying hardware are exposed as APIs for the user-level threads. MANTIS supports preemptive multithreading by assigning priorities to threads, thereby allowing the interleaving of tasks and avoiding delays. Long-running threads can be preempted by short-running threads. Simultaneous execution of these threads is achieved by context switching. When execution of a thread is suspended, all its current states are stored in its own stack and later retrieved to resume execution. Every thread has an entry in a thread table managed by the kernel. Its size is fixed, hence only a predefined number of user-level threads can be created. The other main features of the OS include a dynamic reprogramming mechanism for deployed sensor nodes, a remote debugging mechanism and an x86-based prototype platform. Dynamic reprogramming options are, the wireless re-flashing of the entire OS, the re-programming of single threads and changing the variables of a thread. The wireless re-flashing of the OS and reprograming of a single thread is work-in-progress. A command server is used for remote debugging. The sensor nodes run the client part of the command server. Any user can login to the sensor node and modify its setting, run or stop running threads or restart them. The authors implemented several demanding tasks with MANTIS on MICA2 nodes, including AES and RC5 encryption algorithms, compression/decompression algorithms using arithmetic code, and a 64-bit FFT algorithm. These tasks took low execution time in MANTIS. Normally the concurrent execution of threads leads to context switching overhead and the need for additional stack space. In MANTIS, it was found that while context switching does not incur much performance loss, a stack estimation tool would be helpful.

MANTIS is an interesting option for node-level virtualization, as it is completely thread-based and easier to program without having to manage low-level details of stack/memory. The time-sliced multithreading approach makes it possible to run application tasks simultaneously without using a run-to-completion model. The application threads are coded in C and are independent of the OS. Although MANTIS support dynamic reprogramming but it has not been fully explained in the paper. Currently it is not clear whether the work on MANTIS is underway or not as the project page [31] has quite old information.

The performance results presented in [29] are very limited. No comparison is provided in against other competing solutions. The execution times of some complex tasks (compression/decompression and RC5 and AES encryption) and power consumption using MICA-2 platform are presented.

LiteOS [32] is a Unix-like OS designed for sensor nodes. It provides rich features, such as a hierarchical file system, a command shell that works wirelessly, kernel support for dynamic execution of multi-threaded applications, debugging support and software updates. LiteOS maps a WSN as a UNIX-like file system where different commands can be executed by the user in familiar UNIX-like manner. There are three components: *i)* LiteShell, *ii)* LiteFS and *iii)* Kernel. LiteShell is a command shell that resides in a base station and is used to communicate with sensor nodes to execute file, process, debugging, environment and device related commands. Within the wireless range, sensor nodes can be mounted by LiteShell, similar to how a USB is connected to a computer. However, this process cannot be achieved via the Internet or by multi-hop communication. The sensor nodes do not maintain any state regarding LiteShell and simply respond to the commands.

LiteFS is a hierarchical file system partitioned into three modules that use RAM, EEPROM and Flash memory, respectively. The RAM holds the open files, and their allocation and data information is in EEPROM and Flash memory, respectively. EEPROM holds the hierarchical directory information and the actual data is stored in Flash memory. The LiteOS programming model supports both event-based and thread-based approaches. The scheduling mechanism is also hybrid and supports priority-based and round-robin based scheduling. User applications are multithread-based, and concurrent threads do not have memory conflicts because there is no memory sharing between them. Overall, LiteOS's architecture is inspired by UNIX and works in a distributed manner. The memory consumption of LiteOS applications is larger than that of TinyOS because LiteOS applications are multithreaded whereas TinyOS applications are singe threaded.

LiteOS offers a flexible approach to implement node-level virtualization. It uses a hybrid programming model hybrid that allows the concurrent execution of application threads and handles events through a call-back mechanism. The application tasks can be programmed in C language. Installing and running application tasks is very simple and can be accomplished by dynamically copying user applications. Another advantage of LiteOS is its separation between applications and the OS through callgates. Callgates are pointers and act as application access points to they can access system software and resources. This means that new applications can be simply loaded on a sensor node without reprogramming the sensor node from scratch.

The performance results of LiteShell show the average response time of commands sent using the LiteShell. The average delay of common network commands is under 500ms. The delay to send file in the network using copy command depends on the file size. The delay for 4KB file copy is around 3 seconds to 7.5 seconds for single-hop and two-hop transfer respectively. The length of source code is compared against TinyOS and it is found that the same application can be written in LiteOS using few lines than TinyOS, however because of multi-threading support LiteOS applications take more memory than TinyOS counterparts.

PAVENET [33] OS is a thread-based OS designed to exclusively handle the issues related to the preemption of multithreaded application tasks. However, PAVENET has one major drawback – its non-portability. It only works with PIC18 microchip, and unlike other sensor OSs it cannot be used on other hardware platforms such as MICAZ. Two types of multithreading are provided: preemptive and cooperative. The former is used for real-time tasks (e.g. radio access, sensor sampling) and the latter for best-effort tasks (e.g. routing). PAVENET makes three contributions that deal with the issues of preemption overhead and stack/memory space management; it offers a real-time task scheduler, a best-effort task scheduler and a wireless communication stack to abstract lower layers. To mitigate the effects of switching overheads, the PIC18 chip's functions are used for a real-time task scheduler. One of the functions is the fast return stack that automatically saves the context of a task. The best-effort task scheduler makes use of cooperative task switching to avoid stack/memory issues. The wireless communication stack includes MAC, network and socket layers between the physical and application layers. A buffer is shared by the MAC, network and socket layers to handle the data flow. Tasks with equal priority are grouped together and executed as single task, which leads to code size that is smaller than that of TinyOS. The average clock cycles required to execute an application are better than those required for TinyOS. The support for multithreading means that for complex tasks, PAVENET uses more RAM and ROM than TinyOS.

For WSN virtualization, PAVENET provides a thread-based programming model and uses C language. It is possible to program multithreaded applications with varying priority levels, but their execution will be sequential and not simultaneous because time-sliced execution is not provided. There is also no separation of application tasks from the OS. The main drawback of PAVENET is its lack of portability, although it is an interesting approach that shows how a better CPU design can lead to an efficient sensor OS.

The performance results of PAVENET show that it uses more RAM than TinyOS for sample applications. The execution times of sample applications is comparable to TinyOS. The task switching overhead is found to be 5 times less than MANTIS and comparable to TinyOS. Another aspect is the comparison of lines of codes needed to code sample applications in PAVENET and TinyOS. PAVENET uses twice as less as TinyOS (even more for complex applications).

Contiki [34] is by far one of the most popular systems for WSNs, and over the years has grown to become a leading platform for the IoT and low-powered embedded networked systems. It has a kernel based on an event-driven model, but preemptive multithreading is also provided as an option in the form of a library and exposed as an API for applications to call the necessary functions. Preemption is implemented using a timer interrupt. All threads have their own execution stack.

The concept of protothreads [35] was introduced to combine the concepts of event-driven and thread-based approaches. Protothreads borrows the block-wait approach of threads and combines it with the stack-less approach of events. The advantage of protothreads is that they have lower stack requirement than traditional threads and can be preempted, unlike events. Contiki makes it possible for applications and services to be dynamically uploaded/unloaded wirelessly on sensor nodes. This is made possible by incorporating relocation information in the application binary and later performing runtime relocation.

The OS is written in C language and can be ported to many hardware platforms. CPU multiplexing and an event handling mechanism are the two major functionalities provided by the kernel. The rest of the system-related functionalities are provided as system libraries that can be used by applications when needed. There is no hardware abstraction layer and applications can directly utilize the underlying hardware. Since the OS is event-driven, once an event handler is called, it can only be preempted by an interrupt – otherwise it must run to completion. A simple over-the-air protocol is used to dynamically load/unload applications in a WSN. Binary images of the new application code are sent to selected network nodes using point-to-point communication; the remaining sensor nodes receive the application code as broadcast from them. The current version of Contiki includes several features like full IP support [36], including IPv6 [37], CoAP [38], RPL, 6LowPAN, Cooja, a network simulator to test applications on emulated devices before actual deployment, the Coffee flash file system [39] for sensors that have external flash memory, and a command-line shell for debugging applications.

For node-level virtualization, Contiki is one of the better choices available. It supports multiple applications that are independent of the OS and run on top of it. Applications can be programmed in C language and updated/installed without reinstalling the whole OS. It provides a hybrid programming model. With protothreads, it is possible to create efficient multithreaded applications that share a common stack. Contiki supports many different hardware platforms.

The original Contiki paper used in this work does not provide any systematic performance results. However some insights regarding the performance were presents. For example, reprogramming of a sensor node with a new code (6KB size) took around 30 seconds, whereas the reprogramming of 40 nodes with the same code took around 30 minutes. It is found that code size of similar applications in Contiki is larger than TinyOS but smaller than MANTIS.

TinyOS [30] is another notable effort to provide OS solution for sensor nodes. It is an application-specific, component-based OS based on two characteristics: being event-centric and offering a flexible platform for innovation. It is written in nesC, a dialect of C language, and has a component-based modular design using an event-driven programming model. Three main abstractions are used in TinyOS: commands, events and tasks. Commands are requests to perform a service, events are generated as responses when services are executed, and tasks are functions posted by commands or events for the TinyOS scheduler to execute at a later time. TinyOS components are sets of services, specified by the interfaces that are offered to applications. There are two type of components: modules and configurations. Modules are code snippets written in nesC for calling and implementing commands and events. Configurations connect components through their interfaces. Only components used by the applications are included in the final binary image.

The TOSThreads [40] library was introduced to combine the event-based approach with a thread-based approach, similar to the protothreads in Contiki. Event-based code runs in a kernel thread and user applications run in application threads. Application threads can only run when kernel thread becomes idle. Static optimizations are used during compilation to ensure the removal of any issues in the final code. The OS and the applications are bundled together at compile time in a single file. A component called Deluge [41] is used for over-the-air network-wide reprogramming. The new application code is distributed as composite binaries. Many protocols can be implemented as components. The current version of TinyOS is portable to many hardware platforms.

TinyOS is not the most suitable OS for WSN node-level virtualization. First of all, the programming mode is event-driven and it is often difficult to program event-driven applications. In the context of WSN virtualization, it may not be feasible to bundle applications with the OS at the time of deployment. New application tasks can only be installed by propagating the entire OS image over a virtual machine [42]. TinyOS also has tight coupling between the applications and the OS. The task scheduler in TinyOS is sequential (FIFO based) and executes tasks in run-to-completion mode, meaning a weak form of WSN virtualization.

The performance results of TinyOS highlight important features of the OS. For example, code optimization reduces code size of the programs as much as 60%. The timer component reduces CPU utilization by 38%. The interrupt and task switching also very less time as compared to SenSmart.

*2) Virtual Machine-/Middleware-based Solutions*

Maté [42] is a tiny virtual machine that supports sequential execution and uses a stack-based binary code interpreter. It was designed to work on the early-generation, resource-constrained WSN nodes and it does work on TinyOS. The main purpose of Maté is to enable energy efficient code propagation in WSN with minimal overhead required to re-task sensors. In order to achieve this, application programs are broken into small code capsules and propagated throughout a WSN with a single command. Only predefined applications with predefined instruction sets are possible. There are fixed sets of instructions divided into three classes: basic, s-class and x-class. Basic instructions include arithmetic operations and the activation of sensors/LEDs etc., s-class instructions perform memory access, and x-class instructions perform branch operations. Up to eight user-defined instructions are also allowed. These user-defined instructions need to be fixed when Maté is installed and cannot be changed afterwards. Each program capsule contains up to 24 instructions. Larger programs consist of multiple capsules. The instructions in the capsules are executed in sequence until the halt instruction is reached. New application code is propagated in the network in the form of code capsules, using a viral code distribution scheme. Each capsule contains a version number which is used by a sensor node to determine if it needs to install new application code. Network-wide code propagation occurs when a sensor node forwards the code capsule to its local neighbors, which in turn forward it to their neighbors. Maté maintains two stacks, one for normal instructions and the other for instructions that control the program flow. When an instruction is under execution, a new instruction cannot be executed. This allows for simpler programming options. Maté incurs the cost of byte code interpretations before instructions can be executed. Propagating an 8-byte code to an entire network of 42 sensor nodes required around two minutes.

Regarding node-level virtualization, Maté supports the sequential execution of tasks and tries to address the main drawback of the original TinyOS implementation. New application code can be injected without replacing the OS on a sensor node. However, applications are still tightly coupled. Maté is more suitable for simple event-driven networks where it is possible to define events and their outcomes. To end on a positive attribute, Maté does provide a simple mechanism to automatically reprogram a WSN using code capsules.

The performance results of Maté are collected by implanting an ad-hoc routing protocol which is also implemented in standard TinyOS release with Maté. The implementation of simple operations (such as AND, rand, sense, sendr) take more CPU cycles than native TinyOS, worst-case taking 33 times more CPU cycles and best case taking 1.03 times. A setup of 42 sensor nodes (in a grid pattern) is used to see the propagation of code using Maté. It is found that Maté takes little over 120 seconds to reprogram all sensor nodes with the new code. Overall Maté incurs overhead because its each instruction is executed as a TinyOS task.

VMSTAR [43] is a Java-based software framework for building application-specific virtual machines. It also allows for the updating of WSN applications as well as the OS itself. VMSTAR provides a rich programming interface that allows developers to develop new applications which can be portable to a variety of hardware platforms. VMSTAR generates compact code files rather than regular Java class files. It supports both the sequential and simultaneous execution of thread-based applications. The framework is comprised of three parts: a component language called BOTS [44], a composition tool and an updating mechanism. The component language is used to specify software systems. The composition tool selects/composes the required components and determines the dependencies between them to satisfy specific constraints.

The updating mechanism uses an incremental update technique [45] to take actual coding changes rather than structural changes into account in the program code file like change in number of lines. For simple applications, sequential thread execution is supported, but for complex applications requiring input from external events, two event-based programming models are defined. One is the select model, in which an application subscribes to an event, acquires the corresponding event handle and executes it when the event occurs. In the case of multiple events, the respective handling methods are executed sequentially. The second model is known as action listener, in which applications define event handlers by extending the default handler class from the library – they do not register for events. When an event occurs, the registered callback method is invoked. The action listener model allows for the simultaneous execution of threads, but in the paper only the select model is implemented. A base station is used as a repository for application code and as an orchestrator for deployment and update purposes. A native interface is also provided to allow access to the underlying resources of a MICA platform.

For node-level virtualization, VMSTAR does support the concurrent execution of multi-threaded application tasks but the implementation presented only supports single-threaded Java applications. The programming model is thread-based and applications can be coded in Java language, making it easier for developers. Concurrent events can be handled using action listeners. Although VMSTAR discusses the distinction between the user applications and the OS, for the implementation example both are tightly coupled.

The performance results of VMSTAR show that it performs better than Maté but not so well against native TinyOS. For example, its memory consumption is almost double as compared to TinyOS. The same is true for CPU utilization, where VMSTAR sits between TinyOS and Maté.

Squawk [46] is a small Java virtual machine that runs on sensor hardware. Compared to VMSTAR, Squawk does not require an operating system to run; it provides the required functionalities by itself. These include interrupt handling, networking functions, resource management, support for the migration of applications from one SunSpot to another and an authentication mechanism for deployed applications. Applications in Squawk are represented and treated as objects. Since multiple, isolated objects can reside in a virtual machine, concurrent applications can be executed easily. Squawk VM runs on a specific device platform, Sun Small Programmable Object Technology (SunSpot) which has more processing, memory and storage capability than MICA /MICAZ and other WSN platforms. Squawk VM can use many standard Java features, such as garbage collection, exception handling, pointer safety, and thread library. It is written in Java, in compliance with J2ME CLDC [47]. The device drivers and the MAC layer are also written in Java. Squawk VM supports split VM architecture, where the class file loading is performed on a desktop machine to generate its representation file. The representation file is then deployed and executed on SunSpots. The size of these files is much less than standard Java class files. Green threads are used to emulate multi-threaded environments. The threads are managed, executed and scheduled in user space. An application's status, including its temporary state, can be serialized to a stream for storage. When another Squawk VM, on another SunSpot, reads that stream it can effectively reconstitute the application along with its complete state information. This allows for live-migration of applications from one SunSpot to another. This is quite useful in situations when a SunSpot device is about to run out of battery power.

For node-level virtualization, Squawk VM takes quite a different approach than its competitors. A robust and efficient application isolation mechanism is provided, which allows multiple applications to be represented and treated as Java objects. These objects are instance of the Isolate class and can be started, paused and resumed using available methods. Applications can have multiple threads which are managed by the JVM. The programming model is thread-based and applications can be coded in J2ME. There is also an option for Over-The-Air (OTA) programming which can be used to load, unload, stop and migrate applications on SunSpots.

The performance results of Squawk are presented using some benchmark suits and a math application to measure integer and long computation. For memory footprint, Squawk is compared with KVM for CLDC which shows that Squawk VM with debugging support uses less memory than KVM equivalent. The benchmark suits for Squawk and KVM were run of different sets of ARM platforms with different CPU and memory sizes. The KVM ran on better hardware and hence exhibited better results than Squawk VM. The suits files of applications generated in Squawk have around 37% less size than java class files and JAR files.

Agilla [48] is a mobile agent-based middleware that runs on top of TinyOS and uses a VM engine to sequentially execute multiple application in a round-robin fashion. It uses a mobile agent and tuple-space programming models. The middleware is designed to support self-adaptive applications in WSNs. Application programs are coded as mobile agents that can migrate themselves to other sensor nodes in the WSN in response to changes in the network or in the physical phenomenon that is being monitored. Each sensor node can run several autonomous mobile agents. These mobile agents may perform a strong migration, i.e., transfer application code and its state to another sensor. Weak migration only transfers application code, which means that at its new destination, a migrated mobile agent will restart the application. Agents are injected in the WSN from a base station and propagated one hop at a time. Each mobile agent arrives at a new destination, starts its execution and then migrates to the next-hop sensor node. This process can take quite some time to propagate a new application in the WSN. Each sensor node has a tuple space and a local memory. In a tuple space, data is accessed using pattern-matching techniques. This approach allows mobile agents to be oblivious of each other's memory addresses. Mobile agents have a stack space, a heap and three registers, which are used to store ID of the agent, the program code and condition code. Every agent, including the clones, has a unique ID. The program code register holds the address of the next instruction and the condition code register holds the execution status.

For node-level virtualization, Agilla relies on TinyOS to provide concurrency, and thus mobile agents are executed in a round-robin fashion. However, this is an OS issue, since a multithreaded OS can execute mobile agents in parallel allowing better concurrency. Mobile agents work independently of the TinyOS. The use of tuple-space and locally-stored agent states allows for quick migration, but still much work is left to the programmers to deal with issues such as stalled migration. In a highly dynamic WSN where applications utilize sensor nodes on the fly, such as the IoT, the migration of agents might lead to performance issues. The programming language of Agilla is another difficulty, as the agents are programmed in low-level assembly-like language.

A test-bed of 25 sensor nodes is used to gather the performance results. Agent migration is evaluated by varying number of hops between source and destination sensor nodes. The migration is 99% successful for up to 3 hops but after than it starts decreasing. Also more hops mean more latency, a 5-hop migration can take more than 1.1 second. The latency experienced for remote operations is under 300ms.

The authors in [49] present an integrated system, the UMADE, to promote the utilization of a deployed WSN among multiple contending applications. The main contribution of UMADE is a mechanism to allocate sensor nodes to improve overall Quality of Monitoring (QoM) for the applications. UMADE is implemented on TelosB motes and uses Agilla VM on top of TinyOS. The proposed systems consist of several components such as, specification of QoM attributes, application deployment and relocation of applications to deal with the network changes, as well as QoM-aware application allocation algorithm. QoM attributes are specified by variance reduction and detection probability attributes. A variance reduction QoM attribute exploits the correlation of sensor readings using probabilistic methods to predict sensor readings. For the detection probability QoM attribute, a stochastic model is used to find the probability of an event's detection by a group of sensor nodes. It is not clear from the paper whether QoM attributes can only be specified before the deployment of UMADE or if it is an evolving process. A simple greedy heuristic is used in a QoM-aware application allocation algorithm to maximize the overall WSN utility. Applications are deployed using an application allocation engine and an application deployment engine. The allocation engine runs in a base station and uses an allocation algorithm to find the suitable sensor nodes for an application. The deployment engine, present in both the base station and the sensor node, is used to wirelessly send the sensor application to the chosen sensor nodes. The applications run concurrently in the Agilla VM. Both preemptive and non-preemptive allocation is used to deal with network dynamics and sensor node failure. In preemptive allocation existing applications are relocated to new sensor nodes to increase the overall utility, whereas in non-preemptive allocation no application is relocated to new sensor nodes. The base station side code is written in Java and the sensor node code is written in nesC.

UMADE uses Agilla VM for node-level virtualization. Agilla VM is extended to provide dynamic memory management for concurrent applications. UMADE has event-driven programming model and uses nesC language to code application tasks.

Application specific results are presented in the paper (i.e. application that are implemented for evaluation purposes). For example, an increase in weight of a temperature monitoring application resulted in increase in its utility by 60%. The time to execute multiple application over a set of nodes increases linearly. Since UMADE uses Agilla over TinyOS its performance is highly dependent on those two solutions.

A macro-programming framework, Nano-CF, for the in-network programming and execution of multiple applications over a deployed WSN is presented in [50]. Nano-CF runs over the Nano-RK operating system [51] and allows several applications to utilize a common WSN infrastructure. Using Rate-Harmonized Scheduling (RHS) [52], Nano-CF realizes the coordinated delivery of data packets from multiple application tasks that run on sensor nodes. RHS also allows for data aggregation and ensures that small data packets are combined together before being sent to their respective applications. Nano-CF is a three-layer architecture consisting of a Coordinated Programming Environment (CPE) layer, an integration layer and a runtime layer. The CPE layer is present at the user/programmer side and allows them to write application programs in the Nano-Coordination Language (Nano-CL). Nano-CL is descriptive language with a C-like syntax. Its programs have two sections: service descriptor and job descriptor. The service descriptor section has tasks that are executed by the sensor nodes, as services. The job descriptor section has multiple services along with a set of nodes which will execute them. The programmer has to specify the timing and the periodic rate at which the services (tasks) will be executed at each sensor node. The program code is parsed to byte-code and sent to the sensor nodes by a dispatcher module in the CPE layer. The integration layer is also responsible for handling the data and control packets. It consists of a sender module in the gateway and a receiver module in the sensor nodes to deliver the application task in byte-code. The runtime layer resides in each sensor node and consists of a code interpreter module which translates the received task byte-code for the underlying Nano-RK OS. It also provides routing functionality using DSR protocol. A data aggregation module collects aggregated data from the sensor nodes and sends it to the user application using RHS. The proposed architecture is evaluated using a university campus multi-application sensing test-bed called sensor Andrew [53].

Nano-CF makes several contributions to node-level virtualization. It allows independent application developers to write application tasks for a common WSN infrastructure. Each application task runs independently and is not coupled with the sensor OS. The proposed framework is suitable for data collection applications and for sensor nodes that have multiple on-board sensors. The programming model is event-driven and applications are programmed using their descriptive language, Nano-CL.

The performance results of the solution cover the energy and overhead of code interpreter. Using RHS allows energy savings especially using multiple applications since packets are aggregating first and then transmitted. However, the packet size has an impact on this because bigger packets means they cannot be aggregated due to size issues. When code interpreter is used, the extra-overhead is around 55%.

### B. Network-level Virtualization

We group the network-level virtualization approaches under two umbrellas: virtual network/overlay-based solutions and cluster-based solutions. Virtual network/overlay-based solutions utilize the concept of virtual networks and application overlays to achieve network-level virtualization. Virtual network/overlay are logical networks created on top of physical network(s). In cluster-based solutions, the nodes in a physical network are grouped to work together in connected groups, i.e. clusters. Unlike virtual network/overlays, clustering is more like the physical partitioning of the network where one part of the network is used to one application and another part is used by a different application. Nodes inside a cluster have specific roles, such as cluster-head and cluster-member. Typically cluster-based solutions in WSNs are used to monitor dynamic events.

Fig. 4 shows the network-level virtualization types while Table II illustrates the characteristics of the existing work dealing with node-level virtualization.

#### 1) Virtual Network/Overlay-based Solutions

The work in [9] uses overlays to create application-specific virtual networks on top of the deployed WSN. The overlay is used to allow data exchange between sensor nodes in different administrative domains. This work is more suitable for situations where it is difficult to bundle applications during the deployment of a WSN. A three-layer architecture is presented to allow multiple end-user applications to utilize sensor nodes concurrently. The bottom layer has new-generation sensor nodes like Java SunSpots, as well as older and less capable ones. In order to allow older and less capable sensor nodes to participate in overlays, another entity called Gates-to-Overlay (GTO) nodes is incorporated. The functionality of these GTO nodes can be implemented in gateways and sink nodes, as well as more powerful sensor nodes. The middle layer abstracts the simultaneous tasks executed by the physical sensors as virtual sensors. This is the basic assumption of the work, that the sensor nodes are capable of executing multiple application tasks concurrently. The top layer consists of applications implemented as overlays. These independent applications utilize the data sent by their respective tasks running on the sensor nodes. Each application has an independent overlay with virtual sensors as members of that overlay. This logical grouping allows data exchange even when sensors are physically located in different administrative domains. The architecture has separate paths for data and control messages. A fire monitoring scenario is used as an example, in which the sensor nodes in private homes are used to monitor the progress of fire eruption using a fire contour algorithm. Since sensor nodes are in private homes they cannot send data to each other directly. An overlay network is created to facilitate such data exchange and execute the fire contour algorithm. The authors assume the prior publication of sensor nodes to a registry which the end-user applications use to select the required sensors. The paper does not provide any implementation details. However, certain protocols are suggested for data, control interfaces and for overlays.

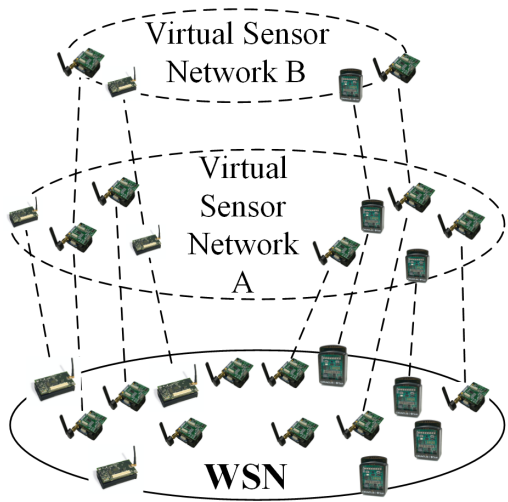

a) Virtual network-based solutions

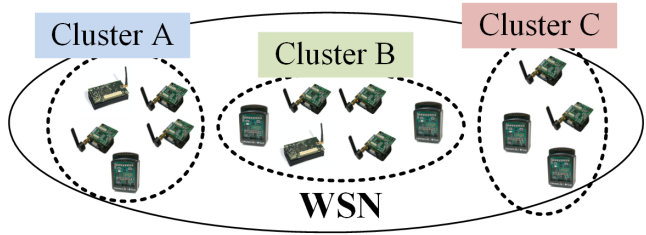

b) Cluster-based solutions

Fig 4: Network-level virtualization solutions

For network-level virtualization this work makes use of application-specific overlays to provide a robust and efficient mechanism for sensors to communicate. There have been some efforts to utilize DHT overlays in WSNs e.g., [54], [55], [56] and [57]. Each sensor can be part of several overlays at the same time and can execute their tasks. In the absence of any implementation details, it is difficult to determine the effectiveness of this solution, but it is quite relevant to IoT, where WSNs will be utilized by different users to provide new applications and services that were not envisioned during their initial deployment. Even geographically dispersed WSNs can be combined to provide data for new applications.

No performance results are presented in this work.

The work in [58] discusses the "Managed Ecosystems of Networked Objects" (MENO) concept, with its broader scope to connect sensor nodes as well as other IP-smart objects to the Internet for end-to-end communication without the use of traditional gateway-based approaches. The idea behind MENO is to create a virtual network on top of physical networks and thereby allow communication with different types of devices, including sensor nodes. Within each virtual network, end-to-end communication is possible using different protocols. Once end-to-end communication is enabled, it becomes possible for application developers to write new applications that utilize sensors, actuators and other devices. This work is still at the conceptual level, without any implementation details or results. It appears to be on track to use a clean-slate approach to integrate the physical world with the Internet in a seamless way. Some motivational scenarios are presented to make a case for integrating WSNs to the Internet.

The concept utilized by MENO is used to develop the Internet of Things Virtual Network (IoT-VN) [59]. That study presents some implementation details by applying the concept of the IoT-VN to constrained and non-constrained environments. For constrained environments, the IDRA framework [60] is used to implement neighbor detection and a tunneling mechanism to create virtual links between the members of the virtual network. For non-constrained environments, the Click Router [61] is used, a C++ based framework capable of realizing network packet processing functionality. Routing the data over virtual links is accomplished by means of the AODV protocol. They have extended the AODV header to include IoT-VN ID header and a network header. A simple ping application implements basic request and reply messages to demonstrate data exchange inside a virtual network.

For network-level virtualization, the work in [58] and [59] uses the concept of virtual links built over either layer 3 or layer 2 in traditional networks, and over IEEE 802.15.4 in WSNs. Not much detail about the actual protocols is provided, but these researchers do mention some motivational scenarios to open up WSN deployments and connect them to the Internet. Overall, the focus here is on connecting different devices (resource-constrained and non-resource constrained) together and allowing end-to-end communication for the deployment of new applications and services.

The work in [58] does not provide any performance results, however in [59] presents early results using a simple two sensor test-bed setup. Round trip times of a ping command are shown which was sent from one sensor to another. Overall the results do not give much insight in to the solution.

An embedded agent-based approach is presented in [62] to create and maintain virtual sensor networks. This agent-based solution is built on top of Java SunSpot devices, as they offer Java programming support and are easier to program. The authors first provide an analysis of the layered approach normally used to create and maintain a Virtual Sensor Network (VSN). In this layered approach a new VSN layer is introduced to create and maintain a VSN, but this approach is not flexible when the sensor nodes' sleep and wake patterns are taken into account. A sensor node that is part of more than one VSN at a time cannot sleep abruptly without first coordinating with other sensor nodes to inform them about its unavailability. Since the layers in sensor nodes are tightly coupled and cannot be changed without affecting the other layers, an agent-based solution is proposed in this work. Agent Factory Micro Edition (AFME) [63] library is used to create agents on a sensor node. Each agent resides on a sensor node and is responsible for creating and maintaining a VSN, as well as for communicating with the agents working for the same VSN on other sensor nodes. These agents can communicate with each other to optimize performance. AFME allows communication between agents for easy message exchange. AFME also allows the migration and cloning of agents in the network, which makes it easy for new sensor nodes to join a VSN. Using the agent-based approach has obvious benefits, not least because a sleep broker can make intelligent decision about the sleep and wake duration of sensor nodes.

For network-level virtualization the work in [62] considers independent VSNs created over a WSN for different applications. To create such VSNs, mobile agents create a virtual topology linking sensor nodes together for an application. Although the agents are implemented using AFME, there are no details about VSN formation and its operation.

Interestingly the work does not provide any performance results of the agent-based approach instead it present simulated results of layered approach showing their obvious drawbacks.

Pioneering work regarding network-level virtualization was first presented in [19] and extended in [64] and [65]. In [19], a subset of WSN nodes dynamically forms a VSN. Applications with attributes or situations such as being geographically dispersed, using heterogeneous WSN nodes with different capabilities and that monitor dynamic phenomenon are particularly suited to take advantage of VSNs. Each independent subset executing an application is a VSN. In this approach, it is clear that different applications can execute sequentially, due to the dynamic VSN formation by different node subsets. However, the authors do not give any information about how these applications might eventually be executed simultaneously. Two illustrative applications are presented. One is a geographically overlapped application which works in scenarios where heterogeneous WSN nodes are deployed to monitor two different events spread over a large area. Each WSN needs to be deployed without using resource sharing even in those areas where there is no event of interest, to provide communication and routing. With resource sharing however, other WSNs can help, resulting in a more efficient use of resources.

The second application illustrates the concept of monitoring a dynamic event with a subset of WSN nodes. This subset can expand or reduce depending on the dynamics of the event. The work discusses the management issues of these VSNs and describes functions to create VSNs. WSN nodes that are not part of any subset help in the overall WSN operation, with data routing for example, or remain asleep to conserve energy.

For network-level virtualization the authors in [19] present the basic motivation to create VSNs. Example applications are discussed. However, the paper presents high-level details and does not include any technical details, e.g. how to realize these VSNs. The paper provides the basic concept of multiple applications sharing a WSN and using multiple WSNs for new applications without additional deployments.

No performance results are presented in this work.

*2) Cluster-based solutions*

A self-organizing tree-based solution is presented in [64] to facilitate the creation, operation and maintenance of VSNs. When an event has been detected, a dynamic cluster tree is formed, ensuring that nodes will join a VSN to monitor the event in a reactive manner. In this approach the sequential execution of applications is possible, since VSNs are formed dynamically, but it is not clear if (or how) it is supported by the WSN nodes. This approach uses cluster heads and child cluster heads inside VSNs to carry out different functions. This structural organization provides logical connectivity among WSN nodes and ensures that two different notifications of the same event are detected and treated as one; meaning that no event in the deployed WSN remains unknown. Once an event is detected, a dynamic cluster tree is formed by exchanging VSN formation messages.

VSNs provides unicast, broadcast and multicast communication. For unicast communication, a hierarchical addressing scheme like DNS is used while broadcast and multicast communication use a list. This list is used by each cluster head to keep track of the child cluster heads it serves. A new hierarchical clustering algorithm is proposed to create VSNs. A simulation-based performance analysis of the proposed algorithm is presented using a custom-built simulator in C language. However, advanced VSN functions like the merging and splitting of VSNs are not implemented.

A cluster tree mechanism is used to group the sensor nodes that work for an application, as a way to realize network-level virtualization. This work is an extension of the work in [19]. Dynamic trees are formed and communication between the sensor nodes is also supported. There is no discussion about the physical implementation of this proposed scheme.

TABLE II
CHARACTERISTICS OF NETWORK-LEVEL VIRTUALIZATION SOLUTIONS

| Solution (year) | Network Formation Mechanism | Protocols or Algorithms used | Results in the Paper are based on |
|---|---|---|---|
| Khan et al. (2013) | Application Overlays | JXTA | Not discussed |
| IoT-VN (2012) | Virtual Links | AODV | Implementation |
| MENO (2011) | Virtual Links | Not discussed | Not discussed |
| Taynan et al. (2008) | Virtual Network using Embedded Agents | Coverage Configuration Protocol (CCP) and Interpolation-based Redundancy Identification and Sensor Hibernation (IRISH) | Simulation |
| Jayasumana et al. (2007) | Virtual Links | Not discussed | Not discussed |
| Dilum et al. (2008) | Cluster Tree | Hop-ahead Hierarchical Clustering (HHC) | Simulation |
| Han et al. (2008) | Cluster Tree | PHenomena AwaRE clustering in wireless sensor networks (PHRE) and DRAGON | Implementation |

For performance results a discrete-event simulator is used. Three scenarios are implemented to detect events in different regions and use sensor nodes to monitor them. The results show a linear increase in number of hops similar to the increase in sensor nodes monitoring the event. When an event occurs, with source and destination node in the same region, more unicast messages are exchanged but these messages are not affected by the network size. On the other hand, when an event occurs in another region more multicast messages are exchanged and are affected by network size.

A proof-of-concept study that monitors an underground plume is presented in [65]. The proof-of-concept is based on a single application, and so it is difficult to find a link with sequential or simultaneous execution. The authors also discuss a phenomena-aware clustering algorithm to create and maintain VSNs. Using this algorithm, clusters are comprised of groups of WSN nodes that are close to dynamic phenomenon and report on it frequently throughout their lifetimes. With these reports, the algorithm is able to select those WSN nodes which are relevant for clusters and that are close to the dynamic phenomenon, allowing less-relevant WSN nodes to save their energy for other applications. This technique considerably reduces the required data reporting since only relevant data is sent. As the deployed WSN is event-based and not always on, sudden bursts of data are avoided whenever an event of interest occurs. The algorithm is also resilient to WSN node and link failures. To adapt to the dynamics of an event, i.e., a merger or a split, another algorithm, called DRAGON, is presented. When an event is detected, DRAGON ensures its location is found and used as a reference point to track its movement. Sensor readings and the relative positions of WSN nodes are then used to make decisions about whether two events should logically remain distinct or be merged into a single event.

For network-level virtualization this work is based on [19] and [64]. The proof-of-concept prototype is used to demonstrate the viability of the concepts presented in earlier papers, however only one application is demonstrated.

There are not much performance results of the prototype except that the sensors were able to track a plume similar to the conductivity probes.

*C. Hybrid Solution*

Hybrid solutions combine both node- and network-level virtualization mechanisms. We group the Hybrid solutions under three types: middleware and cluster-based solutions, middleware and virtual network/overlay-based solutions and virtual machine and dynamic grouping-based solutions.

In middleware and cluster-based solutions, a middleware handles node-level virtualization, while network-level virtualization is achieved by grouping sensor nodes into clusters. In middleware and virtual network/overlay-based solutions a middleware handles node-level virtualization while network-level virtualization is achieved using virtual network/overlays. In virtual machine and dynamic grouping-based solutions, node-level virtualization is achieved using a virtual machine, and a tailored, sensor node grouping scheme is used for network-level virtualization.

Fig. 5 shows the hybrid virtualization solution while Table III shows the characteristics of hybrid solutions.

*1) Middleware and Cluster-based Solutions*

In [66], a middleware solution, Sensomax, for Java SunSpot [67] devices is presented. Sensomax follows a components-based approach and provides several operational paradigms such as data-driven, event driven, time-driven and query-driven, to offer more flexibility. The main contributions of Sensomax are support for multi-tasking, dynamic task modification and re-programming at runtime. At node-level, user applications are coded as application-specific agents. Concurrency is implemented using a main Monolithic Kernel, abstracting the sensor resources. Applications act as Microkernels running atop the Monolithic Kernel and access underlying resources in a uniform way. When an application starts its execution in a sensor node, its corresponding agent is loaded to an execution space and queued for execution. A resource-algorithm is said to be used for allocating resources to multiple agents in the execution space. However, no details of such allocation algorithms are discussed. Application agents can be data-driven, event-drive, time-driven, query-driven or hybrid models.

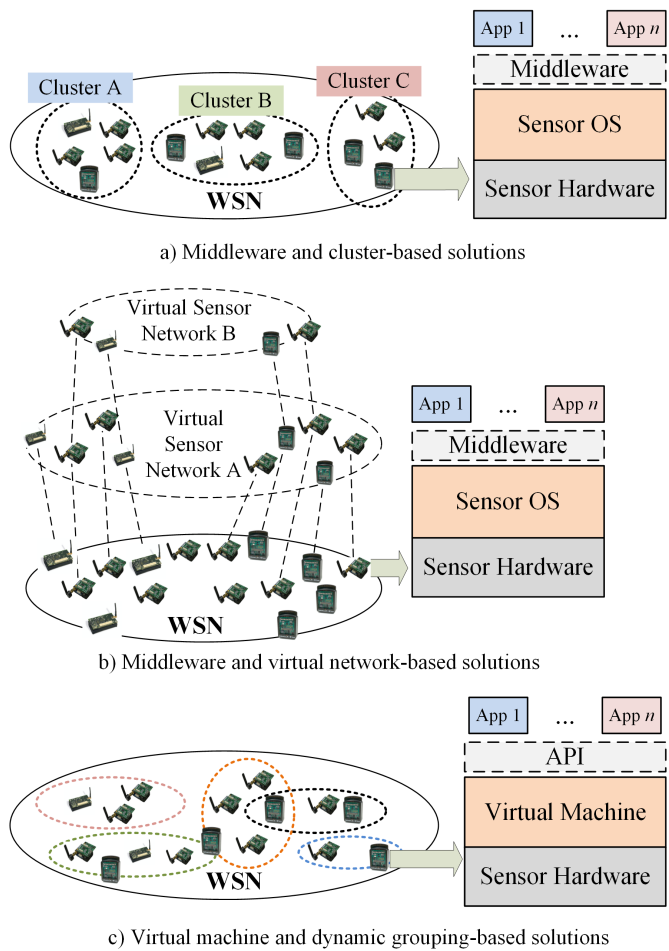

Fig 5: Hybrid virtualization solutions

At the network level, the deployed WSN is divided into multiple clusters consisting of sensor nodes. Each cluster is dedicated to a single or multiple applications and treated as a single entity by the application programmers. The applications can span over multiple clusters by running application-specific agents in each cluster. Each cluster consists of a sensor node acting as the cluster-head and several sensor nodes acting as cluster members. Sensor nodes can have dual roles, i.e., a sensor node can act as cluster-head for an application while at the same time it can be a cluster member for a different application. Such roles depend on the application agents residing in a sensor node. The agent-based approach is used for network-level communication in Sensomax. The global agents enable different network entities to communicate with each other. The local agents are used for intra-cluster communication, allowing the cluster-heads to communicate with their cluster-members and vice-versa. The system agents are used by the base-station to send configuration instructions to cluster members via cluster heads. The system agents are used to reprogram or update sensor nodes on the fly. The WSN resources are divided into three main classes: global, local and system resources. Global resources include sensors, actuators and processes that are shared among different network entities. Local resources include resources found inside a cluster and can only be shared between members of that particular cluster. System resources include items such as system properties where resource states are defined. A one-hop broadcasting of agents is used to propagate application-specific agents in the WSN. Java SunSpot devices are used for implementation and simulation-based results are presented.

For node-level virtualization, Sensomax uses Java SunSpot devices and exploits their ability to run concurrent application tasks. Each user application is programmed as an agent, and multiple agents can reside on a single sensor node. Agents are submitted via a base station and propagated into the WSN using a one-hop broadcast. The network-level virtualization uses the clusters concept. The WSN is divided into multiple clusters, each with its own cluster head. Different types of communication modes are provided to enable communication between different network entities.

The performance results are collected by means of a test-bed consisting of 12 sensor nodes and a simulator. The processing time of each agent is found to be around 200ms when the sensor node is executing 30 concurrent applications. The simulation results follow the same trend. The sample applications are lightweight that report temperature and light level with various conditions. The dynamic update processing time is under 100ms for the same number of applications.

The work in [68] presents a multi-set architectural model to allow the execution of multiple applications over a deployed WSN. This work is based on the concept of agents, similar to Agilla. The agents are not application-specific, instead they are used to control the node- and network-level functionality. The overall design goal is the ability to run multiple applications in a pre-defined execution order and to be able to adjust their functional parameters. A configuration agent (C-Agent) is used to modify the functional parameters of an application running on a sensor node, e.g., to change its sampling interval. The C-Agent is first propagated in the WSN from the base station to the cluster-heads and then from cluster-heads to the sensor nodes in their clusters. Before the deployment of a WSN, the applications and their order of execution are defined. This step limits flexibility, as new applications cannot simultaneously use the deployed WSN. At node-level, TinyOS is used to provide concurrent execution of application tasks on a sensor node using a middleware that runs on top of TinyOS. The solution inherits the drawbacks of TinyOS; making applications to be executed in their predefined order.

At the network-level, the scoping building block concept [69] is used to divide a WSN into subsets. Within these subsets, nodes can be grouped as clusters according to the application requirements. Each subset is dedicated to execute only one application, hence a WSN with $n$ subsets will execute $n$ number of applications. The role of cluster-head is performed by powerful sensor nodes, so there is no selection of cluster-heads on the fly. When the WSN is deployed initially, only one application begins its execution, according to a pre-defined sequence. The sensors in other subsets sleep to conserve their energy until it is their turn to execute their application. A switching agent (S-Agent) is used to switch from one application to another by putting awake sensor nodes into sleep mode and vice-versa. There is no information about how these agents are propagated.

For node-level virtualization, the solution works similar to the TinyOS and provides a weak form of virtualization. Pre-defining applications and their execution sequences does not make this solution very attractive. For network level virtualization, the WSN is divided into subsets that have multiple clusters. At any given time the sensor nodes in one subset are active while others sleep to save their energy.

No performance results are presented in this work.

*2) Middleware and Virtual Network/Overlay-based Solutions*

The authors in [70] discuss SenShare, a platform to execute multiple applications over a WSN. This is the first significant effort to tackle the issue of allowing open access WSN deployments running multiple applications concurrently. Two roles, those of WSN infrastructure owners and application developers, are considered. This separation opens up the possibilities for new business models, innovative applications, improved utilization of WSN resources, and flexibility, along with cost benefits. At node-level a hardware abstraction layer (HAL) and a node runtime layer is used in each sensor node to support multiple applications. Each application is a TinyOS program which runs on top of a multi-tasking OS that allows the simultaneous execution of multiple application tasks. The HAL is shared by each application and is used to break the tight coupling between TinyOS applications and the sensor hardware and to allow shared access to the sensor hardware. Each application contains virtual hardware controllers (e.g. access to LEDs, sensors, timers and network I/O) that are linked to all TinyOS application at compile time. When an application requires access to, e.g. a sensor, the corresponding virtual hardware controller passes the request to a runtime layer between the applications and the multi-tasking OS. The runtime layer is OS-specific and all of the TinyOS applications use it to access the sensor hardware. It runs as a separate process inside every sensor node and mediates between the applications and the sensor hardware. The sensor I/O and network I/O are two components in the runtime layer that allow managed access to sensing components and to the network interface, respectively. This access is allowed asynchronously to multiple applications. Each application in SenShare, has a unique ID which is used to manage it. In order to deploy an application, SQL-like commands are used to select the target nodes according to the application's requirement. Afterwards the application's binary code is sent to the selected nodes using a modified version of the Deluge protocol [71]. Once the application is up and running, the virtual topology is formed to provide isolation from other data/control traffic. The WSN is globally synchronized using the TPSN protocol [72].

At the network level, a network-level overlay is created to group WSN nodes that execute similar application, using the Collection Tree Protocol (CTP) [73]. Physically scattered groups executing similar applications can be joined into a single overlay network. CTP is also used to route data and control messages in the WSN. In order to provide isolation between the traffic from multiple applications, each application packet is modified to include the application ID along with sequence number, origin and destination addresses. The runtime layer attaches and removes this information at the source and destination nodes, respectively.

An application could be executed by physically scattered sensor nodes. Linking these scattered sensor nodes (clusters) into a single virtual connected network requires an overlay formation protocol that utilizes the underlying CTP topology to connect clusters together in a virtual connected network. The protocol works by making each sensor node route its packets to the closest cluster.

For node-level virtualization, SenShare implements application tasks as TinyOS programs over a multi-tasking OS. The programming model is similar to TinyOS. Incorporating virtual hardware controllers with the applications makes the solution less flexible, as developers need to be aware of the type of hardware each sensor node has. The runtime layer between the OS and the applications does not expose the sensor hardware to the developers, so they cannot write applications on the fly. For network-level virtualization, SenShare uses the concept of overlays and uses CTP protocol to create independent overlays for applications.

The performance results of this work cover the application isolation penalty and overlay management. With more concurrent applications in a sensor node, it is observed that sampling rate decreases by 28% as compared to a single application sampling the same phenomenon. The CPU utilization also increases linearly and has less impact of the SenShare runtime. The same is observed for memory usage. The extra overlay traffic is found to be decreasing over the period of time to around 10% of the network traffic.

The work in [10] discusses the node- and network-level virtualization of sensor nodes in the context of the VITRO project. The goals of this work are *i)* to design a middleware to act as a bridge between applications and the sensor nodes, and *ii)* to design advanced sensor node architecture. Node-level virtualization is achieved by instantiating various instances of routing and of MAC layers. There is a node virtualization manager (NVM) inside every sensor node which is responsible for managing the available resources and fulfilling the requests to utilize those resources [74]. NVM interacts with each layer to ensure the optimal, secure and energy-efficient utilization of sensor nodes. Each sensor node has a middleware which is responsible for its discovery and the service it provides. This middleware sits on top of the network layer, which is responsible for the routing data. The network layer uses routing protocols that can support multiple routing instances. A trust-aware routing protocol [75] is used to route the data, and delay-tolerant network mechanisms are suggested to counter the connectivity issues. For each application, a newly configured MAC layer is instantiated.

TABLE III
CHARACTERISTICS OF HYBRID SOLUTIONS

| Solution (year) | Programming Model | Programming Language | Separation between OS & Application Tasks | Real-time Applications | Protocols or Algorithms used | Network Formation Mechanism | Results in the Paper are based on |
|---|---|---|---|---|---|---|---|
| Sensormax (2013) | Thread-based | Java | Yes | No | Market-based resource allocation algorithms | Clusters | Simulation & Implementation |
| ShenShare (2012) | Event-driven | nesC | Yes | Yes | Collection Tree Protocol (CTP) | Network-level Overlays | Implementation |
| VITRO (2012) | Not discussed | Not discussed | Yes | No | Not discussed | Not discussed | Not discussed |
| Majeed et al., (2010) | Event-driven | nesC | No | No | Not discussed | Clusters | Not discussed |
| Melete (2006) | Event-driven | TinyScript | No | Yes | Trickle | Connected Graph | Simulations & Implementation |

A reference architecture is presented at the network level, consisting of several autonomous WSN domains. Each of these domains is connected to VITRO service providers through a gateway node. The gateway node plays a major role in providing network-level virtualization. It consists of modules that help in the creation and management of VSNs. The gateway node uses several registries to create and manage a VSN. In VITRO, only gateway nodes can be part of the VSN, which can be realized by creating a routing link between the gateway nodes using protocols such as RPL. Individual sensor nodes can only be part of the VSN, on their own, if they support the functionalities of the gateway node, otherwise they can only join a VSN with the help of a gateway node. Details such as sensor selection and task dissemination are not discussed. A VSN manager is responsible for service negotiation, session establishment and monitoring. Functional architectures of gateway nodes and advanced sensor nodes are also presented, along with the details of the interfaces between system components. No implementation details are discussed and no protocol recommendations are given for interfaces or functions such as service registration or service negotiation.

For node-level virtualization, VITRO relies on advanced sensor nodes that enable the efficient utilization of resources and concurrent access. However, there is no discussion regarding the OS that will provide such functionalities, nor is there any information on the hardware platform in the paper. Most of the details are at the conceptual level; no technical details such as programming model, programming language, and OS are provided. For network-level virtualization, this work only connects already VSN-aware/legacy/proprietary WSNs through a gateway node. The mechanisms for creating a VSN-aware network are not discussed, nor is there any mention of protocols to be used.

No performance results are presented in this work.

*3) Virtual Machine and Dynamic Grouping-based Solution*

Melete [18] provides both node and network level support for the concurrent execution of applications in WSNs. At the node level, Melete supports simultaneous execution by enhancing Maté, supporting the interleaved execution of multiple applications on a single WSN node. Application code images are stored, each with its own dedicated execution space. Applications do not share variables with each other to ensure that an application failure does not affect other applications executing on the same WSN nodes. The number of concurrent applications that can be executed by WSN nodes depends on the available RAM; the implementation in the paper supports up to five applications. Melete uses an event-driven programming model. Another contribution of Melete is that it supports application task code dissemination. Task code dissemination has two main goals. One is to select the sensor nodes which are part of a group, and send new code to them. The second is to reactively send code to the sensor nodes that require it. Both goals allow the task code of the relevant sensor nodes to be sent while discouraging its unnecessary dissemination. Actual code forwarding is done region-wise using multi-hop communication.

At a network level, Melete supports the dynamic grouping of deployed WSN nodes in order to execute multiple applications simultaneously. The supported network topology is a connected graph. It is possible for WSN nodes to be part of more than one logical group at a time. Each logical group is dedicated to a single application, and the implementation supports up to 16 groups coexisting in a WSN. A new application code is disseminated passively between members of the group using the above-mentioned design goals. All WSN nodes maintain the version information of the applications, and advertise it in the group, hence making WSN nodes aware of when to update their application codes. This saves energy by reducing unnecessary communications, but at a cost of the delay incurred. Sensor nodes in a logical group execute a single application at a time, hence each application cannot be influenced by the run-time error of another application. The paper presents extensive simulation-based as well as actual implementation results.

For node-level virtualization, Melete improves on Maté, but since application tasks have their own data and execution space, only a limited number of application tasks can run concurrently. The programming model is based on the event-driven approach of TinyOS. The application programs are written in TinyScript. A dynamic grouping scheme is provided for network-level virtualization. By default, all sensor nodes are members of a parent group, with its code stored in them. How a sensor node will join a new group depends on the task code it is executing. The programmer needs to be aware of the many situations that may arise in the network and program the responses, and this approach is not flexible at all.

The performance results of Melete include mathematical analysis of the impact of parameters on the task code dissemination scheme. The code size and memory consumption of Melete was compared to Maté. The code size of Melete is bigger than Maté even when there was only one application. Similarly Melete exhibits higher memory consumption than Maté. The result pertaining to dynamic grouping show delays in the order of seconds for a motion tracking application in an office setting.

*D. Summary*

Table IV illustrates the evaluation of the existing work based on the requirements identified in section 2.4. We have found several capable node-level virtualization solutions. In the early-generation sensor nodes, the programming model of choice was event-driven, as it was simple to implement, but once its limitations were found, the thread-based approach was used to implement more complex and concurrent tasks in sensor nodes. Of all these works, TinyOS and Contiki have become extremely popular and have good community support. Contiki is now considered as a platform for the IoT [76] and has incorporated many innovative features over the last decade. RIOT [25] is a new work to design a capable OS to run C/C++ applications on heterogeneous sensor platforms.

For network-level virtualization, the early work used the concept of clusters but managing clusters itself is quite challenging. The majority of work on cluster-based solutions in WSNs is focused on improving routing, energy efficiency and security. We need solutions that facilitate the creation of application-specific clusters that adapt to the dynamics of the network and of the monitored events. Recently overlay solution are being used for network-level virtualization but it is still largely unexplored territory. We have works like [54] discussing, quite convincingly, that it is not '*mission impossible*' to use overlays in WSNs. Most recent research work has focused on providing hybrid solutions for WSN virtualization. A few recently-concluded research projects have addressed WSN virtualization, but their solutions are embryonic and multiple issues remain. For example, some solutions are platform dependent, others are theoretical and at conceptual level.

TABLE IV
SUMMARY OF THE STATE-OF-THE-ART

| Solution | Type | Node-level Virtualization | Network-level Virtualization | Application Priority | Platform Independent | Resource Discovery | Application to Resource-Constrained Nodes | Heterogeneity | Sensor Selection for Application Tasks |
|---|---|---|---|---|---|---|---|---|---|
| SenSmart [23] | OS-based | Yes | No | No | Yes | No | Yes | Yes | No |
| RIOT [25] | OS-based | Yes | No | Yes | Yes | Yes | Yes | Yes | No |
| PAVENET [33] | OS-based | Yes | No | Yes | No | No | Yes | No | No |
| SenSpire [28] | OS-based | Yes | No | Yes | Yes | No | Yes | Yes | No |
| Nano-CF [50] | VM-based | Yes | No | No | Yes | No | Yes | Yes | No |
| UMADE [49] | VM-based | Yes | No | No | No | No | Yes | No | Yes |
| Agilla [48] | VM-based | Yes | No | No | No | Yes | Yes | Yes | No |
| LiteOS [32] | OS-based | Yes | No | Yes | No | No | Yes | Yes | No |
| Squawk VM [46] | VM-based | Yes | No | Yes | Yes | No | No | No | No |
| VMSTAR [43] | VM-based | Yes | No | Yes | No | No | Yes | No | No |
| MANTIS [29] | OS-based | Yes | No | Yes | No | No | Yes | No | No |
| TinyOS [30] | OS-based | Yes | No | No | Yes | No | Yes | Yes | No |
| Contiki [34] | OS-based | Yes | No | Yes | Yes | No | Yes | Yes | No |
| Maté [42] | VM-based | Yes | No | No | No | No | Yes | Yes | No |
| Khan et al. [9] | Overlay-based | No | Yes | Yes | Yes | Offline publication | Yes | No | No |
| MENO [58] | Vm-based | No | Yes | – | – | – | – | – | – |

| Solution | Type | Requirements | | | | | | | |
|---|---|---|---|---|---|---|---|---|---|
| | | Node-level Virtualization | Network-level Virtualization | Application Priority | Platform Independent | Resource Discovery | Application to Resource-Constrained Nodes | Heterogeneity | Sensor Selection for Application Tasks |
| IoT-VN [59] | VM-based | No | Yes | No | Yes | Yes | Yes | No | No |
| Taynan et al [62] | VM-based | No | Yes | No | No | No | No | No | No |
| Jayasumana et al [19] | VM-based | No | Yes | No | Yes | No | Yes | No | No |
| Dilum et al [64] | Cluster-based | No | Yes | No | Yes | No | Yes | No | Yes |
| Han et al [65] | Cluster-based | No | Yes | No | Yes | No | Yes | No | Yes |
| Sensomax [66] | Middleware- and cluster-based | Yes | Yes | No | No | No | No | No | No |
| SenShare [70] | Middleware- and overlay-based | Yes | Yes | Yes | No | Yes | No | No | Yes |
| VITRO [10] | Hypervisor- and VM-based | Yes | Yes | No | Yes | Yes | Yes | No | No |
| Majeed et al. [68] | Middleware- and cluster-based | Yes | Yes | Yes | No | No | Yes | No | Predetermined |
| Melete [18] | VM- and Dynamic grouping-based | Yes | Yes | No | No | Yes | Yes | No | Yes |

### V. WSN VIRTUALIZATION RESEARCH PROJECTS

In this section we introduce some relevant projects that envision the utilization of WSNs by multiple applications. Table V lists these projects and provides their summary based on the following characteristics.

*1) Project Aim*

Provides the holistic aim of the overall project. FRESnel and VITRO are the only two projects that are aimed directly at WSN virtualization. The remaining projects have more extended scopes, such as smart city realization, smart health in the context of IoT, or aim to provide a large-scale test bed for network research.

*2) Project Scope:*

Indicates if a project is a part of academic or industrial research, or is being developed as a multi-partner effort. VITRO, Smart Santander, iCore and Butler are all European FP7 projects involving large consortiums of industrial, telecom and academic partners. FRESnel is a joint project between Cambridge and Oxford Universities, UK.

*3) Virtualization Level:*

Indicates the type of WSN virtualization. FRESnel and VITRO are the two projects that aim to provide both node- and network-level virtualization. CitySense, iCore, Butler and ViSE do not explicitly address WSN virtualization, but they do consider the utilization of sensors by multiple applications.

*4) Virtualization Type:*

The true realization of WSN virtualization does not involve any gateway node managing the virtualization-related tasks; instead, sensor nodes themselves handle such tasks. On the other hand the gateway-based virtualization solutions make WSNs act as capillary networks connected to the Internet or to other networks through a single node. It is important to mention that the presence of a gateway node for communication cannot be ruled out.

*5) Network Devices:*

Another important characteristic of these projects is the type of devices they use in their work. CitySense, Butler and ViSE use high-end devices. While sensors are considered, they are usually connected to high-end PCs/nodes that compliment them for processing, data storage, power supply and connectivity. FRESnel and VITRO utilize the usual/normal sensor nodes, which is more relevant to WSN virtualization.

*6) Evaluation Setup:*

All of the projects discussed here evaluate their contributions using real test bed setups; however the size of these setups varies considerably. For example, the Smart Santander project will use around 20,000 nodes deployed over four European cities, providing a massive platform for research and evaluation purposes. This gigantic setup will also be used by the iCore project. In comparison, ViSE has a test bed of 3 nodes and Fresnel's in-campus test bed has 35.

The ViSE and CitySense projects were not designed to provide solutions for WSN virtualization, but they do incorporate the important virtualization concept, i.e. to allow multiple applications to utilize the deployed WSN infrastructure. The Smart Santander, iCore and Butler projects are aimed to realize the IoT, and consider sensors and devices of different types. VITRO and FRESnel are focused on WSN virtualization, but VITRO provides gateway-based virtualization, which is not a true realization of WSN virtualization. The FRESnel project however, considers the true realization of WSN virtualization, but it provides platform-specific solutions. Overall it is clear that the idea of WSN virtualization is receiving considerable attention, not only from academic quarters but also from major industrial and telecom players.

TABLE V
WSN VIRTUALIZATION-RELATED PROJECTS

| Project (Year) | Project Aim | Project Scope | Virtualization Level | Virtualization Type | Network Devices | Evaluation Setup |
|---|---|---|---|---|---|---|
| CitySense [77] | Provide city-wide test bed for distributed & networking research | Academic research | Network-level | Gateway-based virtualization | Embedded PCs with Linux acting as gateways | 100+ PCs distributed over an urban area |
| FRESnel [78] (2010 - 2012) | Provide a federated WSN framework for multiple applications | Academic research | Node- and Network-level | Sensor node-based virtualization | iMote2 nodes using embedded Linux | 35 iMote2 nodes distributed in an academic building |
| VITRO [74] (2010 - 2013) | Develop architectures, algorithms to provide VSNs. | Academic research + Industry | Node- and Network-level | Gateway-based virtualization | TelosB, IRIS, iSENSE, xbee, TmoteSkye, AdvanticSys kit | Simulations + 5 test bed setups by project partners |
| Smart Santander [79] (2010 - 2013) | Provide a city-wide IoT experimentation platform for smart city applications | Academic research + Industry | Network-level | Gateway-based virtualization | Sensor nodes, IoT devices, RFID tags, GPRS devices | About 20,000 sensors deployed in four European cities |
| iCore [80] (2011 - 2014) | Provide a cognitive framework consisting of virtual objects, composite virtual objects & business perspectives | Academic research + Industry | Abstract representation of sensors | Gateway-based virtualization | Sensors, ICT devices, everyday objects | Will utilize the Smart Santander test bed |
| Butler [81] (2011 - 2014) | Provide a secure, pervasive, energy-efficient & context-aware architecture | Academic research + Industry | Abstract representation of sensors | Gateway-based virtualization | Smart objects, mobile devices and smart servers | Several field-trials and proof-of-concepts are planned |
| ViSE [82] (2008 - 2011) | Provide public access to a WSN test bed using the GENI framework for multiple users | Academic research | Abstract representation of sensors | Gateway-based virtualization | High-end nodes running Linux and acting as gateway nodes | Three nodes deployed in a town near a forested area |

## VI. RESEARCH ISSUES

We identify some important research issues that need to be addressed to provide innovative WSN virtualization solutions.

### 1) Advanced Node-level Virtualization

Node-level virtualization has attracted considerable attention from the research community. In many ways, it is provided as part of the sensor OS. Multi-threaded OSs and application-specific virtual machines (VM), working on top of an OS, can support the concurrent execution of application tasks. As the trend moves towards more powerful IP-WSNs, more efforts are required to virtualize the individual components of sensor nodes, such as MAC and routing layers. The VITRO project has put forth the concept [10], but there are no real implementations to date. PAVENET OS [33] takes advantage of capable hardware to design efficient OSs but is tied to a single platform. To exploit the recent advances in sensor hardware, a fresh approach like RIOT OS [25] can be taken to come up with new and general purpose solutions. Some new solutions provide separation between the sensor OS and the user application tasks but we still need functions like OTA installation/updating of new user tasks without disturbing the existing ones. One possible solution to tackle this issue is to design an abstraction layer that works on top of sensor OS to provide application portability like in [83]. A modular-based approach will work much better since it will be applicable to heterogeneous OSs, programming languages and models.

### 2) Network-level Virtualization

Not much work has been done in the area of network-level virtualization to support multiple applications over a deployed WSN, hence there is a tremendous opportunity to make valuable contributions. Overlay networks can provide an efficient solution as they are robust and can work efficiently without changes in the underlying network. Some solutions exist like those in [54], [56] and [57] do exist, but they are still embryonic in nature and do not consider the requirements of multiple applications utilizing a WSN concurrently. As multiple overlay may need to co-exist, preventing them from interacting with each other in a harmful way remains a challenge. Cluster-based approaches have traditionally been used in WSN's for improving routing, energy-efficiency, management and security. Managing clusters in a virtualized WSN is not trivial, however, cluster-based solutions can be quite useful in scenarios where a deployed WSN is used to monitor dynamic events. These solutions can also be helpful in mobile WSNs, Robotic and Vehicular Ad hoc Networks.

### 3) Discovery and Publication

The discovery and publication of resources and services in WSN is already challenging, but it becomes even more sophisticated in virtualized WSNs. For example, it will be interesting to find whether certain kind of relationships exist between physical and virtual sensors and whether they can be exploited to provide quick publication and discovery solutions. As virtual sensors are created on-demand and destroyed when no longer required, their publication and discovery needs to be efficient, robust, scalable and manageable. Discovery and publication of resources and services on the fly are very important functions, especially in the context of IoT. A P2P based architecture can be a solution like [84] that does not rely on any central mechanism to discover the services. However, no such solution exists for virtualized WSNs. Similarly a service recommendation system can be developed, for virtualized WSNs, which allows context-aware discovery of resources and services. Recent IETF service discovery protocols like CoAP resource discovery [85], [86] and DNS-SD [87] can be used to design efficient discovery and publication solutions in resource-constrained environments. Moreover, new algorithms that adapt to evolving WSN conditions and nodes' mobility or failures are required, to ensure service continuity.

### 4) Service Composition

Service composition using virtual sensor nodes is another important research challenge. In our view, future WSN deployments will involve multiple actors, such as WSN providers, virtual sensor providers, service providers, third-party application/services providers and end-user applications. A cloud-based approach could be a solution [88]. WSN resources could be offered as Infrastructure-as-a-Service (IaaS) and used by Platform-as-a-Service (PaaS) to offer services to end users. In this regard, existing projects like [79], [80] and [81] can be used for inspiration about end user services. Using semantics and ontologies to compose services based on application requirements and the capabilities of sensor nodes can provide improved solutions. It is also important to note that the service composition may use existing or third-party services on the fly. Location and mapping services are typical examples of such services.

### 5) Sensor Node Selection and Task Assignment

The issues of sensor selection and task assignment are very much related to each other. Selecting the right set of sensor nodes according to the temporal and spatial requirements of applications is crucial [21] to improving the overall Quality of Monitoring (QoM) systems. A more detailed task assignment problem formulation and its solutions are presented in detail in [89], but it does not consider the possibility of multiple applications using a single sensor node at the same time. In [90] cost-effective market-based algorithms are used for task allocation and resource management. But the proposed algorithms are OS specific (Sensomax) and require more work to determine their suitability. A QoS-aware task allocation algorithm in [91] brings a new dimension into the sensor node selection while satisfying QoS requirements of multiple applications at the same time. New algorithms that not only consider the QoS requirements of the applications but also take into account the properties of the events being monitored by the sensor nodes are needed to advance in this area.

*6) Application Task Dissemination*

When new applications are being contemplated, it is not unrealistic to assume that a new algorithm or application task will need to be sent for the sensor node(s) to execute. Sending the new task code (or updating an existing one) in a seamless way, with no disruption of existing tasks, is quite a challenge. Much of this will depend on the sensor OS and its ability to install and update user tasks without disturbing the existing ones or requiring the reboot of the sensor node. Another issue is how to get the user input and program it to compile it to generate executable code. In the context of IoT, the user may not have technical expertise to code the required program. There needs to be a clear separation between the WSN infrastructure and the user. This can be achieved by having an entity, like service provider, to allow a user to provide her requirements in an easy manner, e.g. in a web-form. This way only some aspects of the (re)programming a sensor nodes can be exposed to the user. Once the input is gathered, the service provider can send it to the physical WSN provider to generate executable code for the selected sensor node(s) and reprogram them. Such a system will have two benefits: one is that the sensor nodes not able to fulfill a task, due to some reason, can be filtered out. Second, based on previous usage patterns of the user, a recommendation system can be devised that makes use of the historical data to recommend and (re)program the sensor nodes. An alternative approach would be to develop a cloud-based PaaS solution and provide toolkits specifically designed to develop, compile, verify, test and deploy sensor application tasks for different sensor platforms.

*7) Reference Designs and Architectures*

A comprehensive virtualization platform for WSNs is required, one that covers all aspects: data acquisition from the sensors, end-to-end communication (including data management and computation), as well as service composition for end-user applications. Such a platform will allow a deeper and complete search space exploration to find the optimal solution for any given WSN application. Furthermore, this complete framework will ensure that all the relevant aspects can be modeled and evaluated comprehensively. Decentralized architectures are required that will enable robust and objective-based solutions depending on application requirements like time sensitivity, QoS, and QoM. Another important aspect is that most of the existing work focuses on the fixed WSNs but in the context of IoT, we can expect more and more deployments of mobile WSNs and even spontaneous ad hoc WSNs. These ad hoc WSNs will be created when large number of sensors communicate together to provide on-demand services for a certain time period and then cease to exist. Participatory sensing and crowed-based sensing, using smart phones, are two forms of the ad hoc WSNs. There is an early work in this area [92] that aims to utilize external sensors with the smart phones. This is achieved by means of a sensor virtualization module developed for the android platform. Still we require more solutions that focus on mobile, ad hoc WSNs and even hybrid variations.

*8) New Protocols, Algorithms and Simulation Tools*

As mentioned in the introduction, recently WSN virtualization is getting attention from the research community and we're now seeing some new contributions in this area. For example, in [93] a harmonized transmission protocol is presented that combines transmissions from a sensor node when it is being used by multiple concurrent applications. References [94] and [95] put forth a reconfiguration scheme and a management scheme, respectively, to manage concurrent applications over a deployed WSN. It will be a good idea to have a capable simulation tool to analyze and evaluate proposed protocols and solutions, simply because initially it may not be possible to have a sizeable WSN deployment for such purposes. A new simulation tool is presented in [96] which simulates multiple concurrent applications over a WSN. While it is a good start, more effort is required to integrate such support in already well-known and established simulation tools.

*9) WSN Virtualization Business Model & Standardization*

A viable business model is required to allow broader (and more commercial) acceptance of WSN virtualization. This can be accomplished easily if WSN entities are decoupled into distinct roles of WSN providers, virtual sensor providers, service providers and third-party applications/service providers. Allowing third-party applications will allow for the rapid development of applications and solutions, since the existing components will be reusable. Another benefit of such business model is that it will pave the way for standardization activities in this area. In our review of WSN virtualization area we strongly felt the need for harmonization between different protocols, data formats, encoding schemes, and consortium-led efforts such as Sensor Web Enablement (SWE) [97]. Currently these incompatibilities act as major roadblocks for proposing generic and open solutions.

*10) Energy Efficient Solutions*

Energy efficiency will remain a key research area in WSNs, even more so when WSN virtualization is involved. While we can safely predict that future sensor nodes will be more capable and resourceful, energy efficient communications, discovery, routing and applications will still be required. So far the main focus has been on making a sensor node sleep for maximum duration possible so that it utilizes less energy. This strategy has worked reasonably well for simple applications but this trend is not sustainable in emerging IoT paradigm. Energy harvesting mechanisms need to be incorporated with WSN platforms as main or alternative source of energy. This will ensure that sensor nodes have a continuous power supply in addition to their batteries. Example of energy harvesting mechanisms are, use of ambient energy like vibrations or solar energy to generate energy [98]. There is considerable research work in this area [99] but commercial platforms are missing.

*11) Access Control, Authentication, and Accounting*

Another important area is to provide a controlled access to deployed WSN resources. In the context of the IoT, sensors deployed by entities like city administrations will probably allow for public access, but they will still require access control, authentication and authorization. For example, such deployments will also be used for monitoring or security applications along with public applications, hence providing access according to users will be challenging. Another aspect is that it may not be feasible for a single authority to deploy a WSN on a large scale. For areas where WSN deployments are not possible, participatory sensing can be used as an alternative. Motivating private owners to share their deployed sensors and allow remote access is a challenge. Incentives like tax rebates or reduced utility rates need to be devised to encourage voluntary participation. Using a WSN deployment for monetary benefits brings in the accounting issue – how to charge users in accordance with service contracts.

*12) WSN Virtualization Application Scenarios and Test-beds*

Applications from domains such as smart cities, smart health, smart homes, green computing and pervasive computing can potentially use the WSN virtualization concept for cost effective solutions. New trends like mobile WSNs, participatory/crowd-based sensing, cloud-based remote sensing and vehicular networks can also benefit from this concept. The availability of test-bed setups like Smart Santander [79] provides a massive basis for prototyping and evaluation purposes.

## VII. CONCLUSION

We have presented a detailed overview of WSN virtualization, as well as the current state of the art. First we categorized WSN virtualization into node-level, network-level and hybrid virtualization, and explained them. We then provided a critical analysis of the existing state-of-the-art in each category and evaluated them based on a set of requirements derived from the motivating scenarios. Several research projects pertinent to this topic were also presented. We outlined several important research challenges and their possible solutions. WSN virtualization is very much relevant in the context of the IoT, in which small-scale devices, at an unprecedented scale, are expected to provide services to multiple applications concurrently, but we have yet to find a comprehensive solution that meets this challenge.

## REFERENCES

[1] M. A. Feki, F. Kawsar, M. Boussard, and L. Trappeniers, "The Internet of Things: The Next Technological Revolution," *Computer*, vol. 46, no. 2, pp. 24–25, 2013.

[2] I. F. Akyildiz, W. Su, Y. Sankarasubramaniam, and E. Cayirci, "Wireless sensor networks: a survey," *Computer Networks*, vol. 38, no. 4, pp. 393–422, Mar. 2002.

[3] S. Loveland, et al., "Leveraging virtualization to optimize high-availability system configurations," *IBM Systems Journal*, vol. 47, no. 4, pp. 591–604, 2008.

[4] N. M. M. K. Chowdhury and R. Boutaba, "Network virtualization: state of the art and research challenges," *IEEE Communications Magazine*, vol. 47, no. 7, pp. 20–26, Jul. 2009.

[5] Z. J. Chong, et al., "Autonomy for Mobility on Demand," in *Intelligent Autonomous Systems* 12, S. Lee, H. Cho, K.-J. Yoon, and J. Lee, Eds. Springer Berlin Heidelberg, 2013, pp. 671–682.

[6] G. Cardone, A. Cirri, A. Corradi, and L. Foschini, "The participact mobile crowd sensing living lab: The testbed for smart cities," *IEEE Communications Magazine*, vol. 52, no. 10, pp. 78–85, Oct. 2014.

[7] H. Ma, D. Zhao, and P. Yuan, "Opportunities in mobile crowd sensing," *IEEE Communications Magazine*, vol. 52, no. 8, pp. 29–35, Aug. 2014.

[8] C. Perera, A. Zaslavsky, P. Christen, and D. Georgakopoulos, "Sensing as a service model for smart cities supported by Internet of Things," *Trans. Emerging Tel. Tech.*, vol. 25, no. 1, pp. 81–93, Jan. 2014.

[9] I. Khan, F. Belqasmi, R. Glitho, and N. Crespi, "A multi-layer architecture for wireless sensor network virtualization," in *Wireless and Mobile Networking Conference (WMNC), 2013 6th Joint IFIP*, 2013, Dubai, UAE, pp. 1–4.

[10] L. Sarakis, T. Zahariadis, H.-C. Leligou, and M. Dohler, "A framework for service provisioning in virtual sensor networks," *J Wireless Com Network, vol.* 2012, no. 1, pp. 1–19, Dec. 2012.

[11] A. Merentitis, et al., "WSN Trends: Sensor Infrastructure Virtualization as a Driver Towards the Evolution of the Internet of Things," *presented at the UBICOMM 2013, The Seventh International Conference on Mobile Ubiquitous Computing, Systems, Services and Technologies*, Porto, Portugal, 2013, pp. 113–118.

[12] Ramdhany, Rajiv and Coulson, Geoff. "Towards the Coexistence of Divergent Applications on Smart City Sensing Infrastructure" *Proceedings of 4th International Workshop on Networks of Cooperating Objects for Smart Cities 2013 (CONET/UBICITEC 2013)*, Philadelphia, USA, April 8, 2013: pp.26-30

[13] E. Patouni, A. Merentitis, P. Panagiotopoulos, A. Glentis, and N. Alonistioti, "Network Virtualisation Trends: Virtually Anything Is Possible by Connecting the Unconnected," in *Future Networks and Services (SDN4FNS), 2013 IEEE SDN for*, 2013, pp. 1–7.

[14] S. Abdelwahab, B. Hamdaoui, M. Guizani, and A. Rayes, "Enabling Smart Cloud Services Through Remote Sensing: An Internet of Everything Enabler," *IEEE Internet of Things Journal*, vol. 1, no. 3, pp. 276–288, Jun. 2014.

[15] C. Liang and F. R. Yu, "Wireless Network Virtualization: A Survey, Some Research Issues and Challenges," *IEEE Communications Surveys Tutorials*, vol. PP, no. 99, pp. 1–1, 2014.

[16] M. M. Islam, M. M. Hassan, G.-W. Lee, and E.-N. Huh, "A Survey on Virtualization of Wireless Sensor Networks," *Sensors*, vol. 12, no. 2, pp. 2175–2207, Feb. 2012.

[17] M. M. Islam and E.-N. Huh, "Virtualization in Wireless Sensor Network: Challenges and Opportunities," *Journal of Networks*, vol. 7, no. 3, Mar. 2012.

[18] Y. Yu, L. J. Rittle, V. Bhandari, and J. B. LeBrun, "Supporting Concurrent Applications in Wireless Sensor Networks," in *Proceedings of the 4th International Conference on Embedded Networked Sensor Systems*, New York, NY, USA, 2006, pp. 139–152.

[19] A. P. Jayasumana, Q. Han, and T. H. Illangasekare, "Virtual Sensor Networks – A Resource Efficient Approach for Concurrent Applications," in *Fourth International Conference on Information Technology, 2007. ITNG '07*, 2007, pp. 111–115.

[20] M. Ceriotti, et al., "Monitoring Heritage Buildings with Wireless Sensor Networks: The Torre Aquila Deployment," in *Proceedings of the 2009 International Conference on Information Processing in Sensor Networks*, Washington, DC, USA, 2009, pp. 277–288.

[21] X. Wang, J. Wang, Z. Zheng, Y. Xu, and M. Yang, "Service Composition in Service-Oriented Wireless Sensor Networks with Persistent Queries," in *6th IEEE Consumer Communications and Networking Conference, 2009. CCNC 2009*, 2009, pp. 1–5.

[22] Dargie, W., Poellabauer, C., 2010. Fundamentals of Wireless Sensor Networks: Theory and Practice. John Wiley & Sons.

[23] R. Chu, L. Gu, Y. Liu, M. Li, and X. Lu, "SenSmart: Adaptive Stack Management for Multitasking Sensor Networks," *IEEE Transactions on Computers*, vol. 62, no. 1, pp. 137–150, Jan. 2013.

[24] S. Nath, P. B. Gibbons, S. Seshan, and Z. Anderson, "Synopsis Diffusion for Robust Aggregation in Sensor Networks," *ACM Trans. Sen. Netw.*, vol. 4, no. 2, pp. 7:1–7:40, Apr. 2008.

[25] Baccelli, Emmanuel, et al. "RIOT OS: Towards an OS for the Internet of Things." *Proc. of the 32nd IEEE INFOCOM. Poster*. NJ, USA: IEEE Press (2013).

[26] H. Will, K. Schleiser, and J. Schiller, "A real-time kernel for wireless sensor networks employed in rescue scenarios," in *IEEE 34th*

Conference on Local Computer Networks, 2009. LCN 2009*, 2009, pp. 834–841.

[27] E. Baccelli, et al., "OS for the IoT - Goals, Challenges, and Solutions, OS for the IoT - Goals, Challenges, and Solutions," in *Workshop Interdisciplinaire sur la Sécurité Globale (WISG2013)*, Troyes, France, 2013, pp. 1-6.

[28] W. Dong, C. Chen, X. Liu, Y. Liu, J. Bu, and K. Zheng, "SenSpire OS: A Predictable, Flexible, and Efficient Operating System for Wireless Sensor Networks," *IEEE Transactions on Computers*, vol. 60, no. 12, pp. 1788–1801, Dec. 2011.

[29] S. Bhatti, J. et al, "MANTIS OS: An Embedded Multithreaded Operating System for Wireless Micro Sensor Platforms," *Mob. Netw. Appl.*, vol. 10, no. 4, pp. 563–579, Aug. 2005.

[30] P. Levis, et al., "TinyOS: An Operating System for Sensor Networks," in *Ambient Intelligence*, W. Weber, J. M. Rabaey, and E. Aarts, Eds. Springer Berlin Heidelberg, 2005, pp. 115–148.

[31] www.cs.colorado.edu/~rhan/sensornets.html (accessed 27/10/2014)

[32] Q. Cao, T. Abdelzaher, J. Stankovic, and T. He, "The LiteOS Operating System: Towards Unix-Like Abstractions for Wireless Sensor Networks," in *International Conference on Information Processing in Sensor Networks, 2008. IPSN '08*, 2008, pp. 233–244.

[33] S. Saruwatari, M. Suzuki, and H. Morikawa, "PAVENET OS: A Compact Hard Real-Time Operating System for Precise Sampling in Wireless Sensor Networks," *SICE Journal of Control, Measurement, and System Integration*, vol. 5, pp. 24–33, 2012.

[34] A. Dunkels, B. Gronvall, and T. Voigt, "Contiki - a lightweight and flexible operating system for tiny networked sensors," in *29th Annual IEEE International Conference on Local Computer Networks*, 2004, 2004, pp. 455–462.

[35] A. Dunkels, O. Schmidt, T. Voigt, and M. Ali, "Protothreads: Simplifying Event-driven Programming of Memory-constrained Embedded Systems," in *Proceedings of the 4th International Conference on Embedded Networked Sensor Systems*, New York, NY, USA, 2006, pp. 29–42.

[36] D. Yazar and A. Dunkels, "Efficient Application Integration in IP-based Sensor Networks," in *Proceedings of the First ACM Workshop on Embedded Sensing Systems for Energy-Efficiency in Buildings*, New York, NY, USA, 2009, pp. 43–48.

[37] M. Durvy, et al., "Making Sensor Networks IPv6 Ready," in *Proceedings of the 6th ACM Conference on Embedded Network Sensor Systems*, New York, NY, USA, 2008, pp. 421–422.

[38] M. Kovatsch, S. Duquennoy, and A. Dunkels, "A Low-Power CoAP for Contiki," in *2011 IEEE 8th International Conference on Mobile Adhoc and Sensor Systems (MASS)*, 2011, pp. 855–860.

[39] N. Tsiftes, A. Dunkels, Z. He, and T. Voigt, "Enabling Large-scale Storage in Sensor Networks with the Coffee File System," in *Proceedings of the 2009 International Conference on Information Processing in Sensor Networks*, Washington, DC, USA, 2009, pp. 349–360.

[40] K. Klues, et al., "TOSThreads: Thread-safe and Non-invasive Preemption in TinyOS," in *Proceedings of the 7th ACM Conference on Embedded Networked Sensor Systems*, New York, NY, USA, 2009, pp. 127–140.

[41] Hui, Jonathan. "Deluge 2.0-TinyOS network programming." at http://www.cs.berkeley.edu/jwhui/research/deluge/deluge-manual.pdf (2005) (accessed 27/10/2014).

[42] P. Levis and D. Culler: "Maté: A tiny virtual machine for sensor networks." In ASPLOSX: *Proceedings of the 10th International Conference on Architectural Support for Programming Languages and Operating Systems*, San Jose, CA, 2002, pp. 85-95.

[43] J. Koshy and R. Pandey, "VMSTAR: Synthesizing Scalable Runtime Environments for Sensor Networks," in *Proceedings of the 3rd International Conference on Embedded Networked Sensor Systems*, New York, NY, USA, 2005, pp. 243–254.

[44] R. Pandey and J. Wu, "BOTS: A Constraint-based Component System for Synthesizing Scalable Software Systems," in *Proceedings of the 2006 ACM SIGPLAN/SIGBED Conference on Language, Compilers, and Tool Support for Embedded Systems*, New York, NY, USA, 2006, pp. 189–198.

[45] J. Koshy and R. Pandey, "Remote incremental linking for energy-efficient reprogramming of sensor networks," in *Proceedings of the Second European Workshop on Wireless Sensor Networks, 2005*, 2005, pp. 354–365.

[46] D. Simon, et al. "Java™ on the Bare Metal of Wireless Sensor Devices: The Squawk Java Virtual Machine," in *Proceedings of the 2nd International Conference on Virtual Execution Environments*, New York, NY, USA, 2006, pp. 78–88.

[47] Muchow, John W. "Core J2ME technology and MIDP." *Prentice Hall PTR*, 2001.

[48] C.-L. Fok, G.-C. Roman, and C. Lu, "Agilla: A Mobile Agent Middleware for Self-adaptive Wireless Sensor Networks," ACM Trans. Auton. Adapt. Syst., vol. 4, no. 3, pp. 16:1–16:26, Jul. 2009.

[49] S. Bhattacharya, A. Saifullah, C. Lu, and G. Roman, "Multi-Application Deployment in Shared Sensor Networks Based on Quality of Monitoring," in *2010 16th IEEE Real-Time and Embedded Technology and Applications Symposium (RTAS)*, 2010, pp. 259–268.

[50] V. Gupta, et al., "Nano-CF: A coordination framework for macro-programming in Wireless Sensor Networks," in *2011 8th Annual IEEE Communications Society Conference on Sensor, Mesh and Ad Hoc Communications and Networks (SECON)*, 2011, pp. 467–475.

[51] A. Eswaran, A. Rowe, and R. Rajkumar, "Nano-RK: an energy-aware resource-centric RTOS for sensor networks," in *Real-Time Systems Symposium, 2005. RTSS 2005. 26th IEEE International*, 2005, p. 10 pp.–265.

[52] A. Rowe, K. Lakshmanan, H. Zhu, and R. Rajkumar, "Rate-Harmonized Scheduling and Its Applicability to Energy Management," *IEEE Transactions on Industrial Informatics*, vol. 6, no. 3, pp. 265–275, Aug. 2010.

[53] A. Rowe, et al., "Sensor Andrew: Large-scale campus-wide sensing and actuation," *IBM Journal of Research and Development*, vol. 55, no. 1.2, pp. 6:1–6:14, Jan. 2011.

[54] Ali, Muneeb, and Koen Langendoen. "A case for peer-to-peer network overlays in sensor networks." *International Workshop on Wireless Sensor Network Architecture (WWSNA'07)*, 2007, pp.56-61.

[55] G. Fersi, W. Louati, and M. B. Jemaa, "Distributed Hash table-based routing and data management in wireless sensor networks: a survey," *Wireless Netw*, vol. 19, no. 2, pp. 219–236, Feb. 2013.

[56] Luu, Hai Van, and Xueyan Tang. "Constructing rings overlay for robust data collection in wireless sensor networks." *Journal of Network and Computer Applications* 36.5 (2013): 1372-1386.

[57] A. A.-B. Al-Mamou and H. Labiod, "ScatterPastry: An Overlay Routing Using a DHT over Wireless Sensor Networks," in *The 2007 International Conference on Intelligent Pervasive Computing*, 2007. IPC, 2007, pp. 274–279.

[58] J. Hoebeke, et al., "Managed Ecosystems of Networked Objects," *Wireless Pers Commun*, vol. 58, no. 1, pp. 125–143, May 2011.

[59] I. Ishaq, J. Hoebeke, I. Moerman, and P. Demeester, "Internet of Things Virtual Networks: Bringing Network Virtualization to Resource-Constrained Devices," in *2012 IEEE International Conference on Green Computing and Communications (GreenCom)*, 2012, pp. 293–300.

[60] E. De Poorter, E. Troubleyn, I. Moerman, and P. Demeester, "IDRA: A Flexible System Architecture for Next Generation Wireless Sensor Networks," *Wirel. Netw.*, vol. 17, no. 6, pp. 1423–1440, Aug. 2011.

[61] E. Kohler, R. Morris, B. Chen, J. Jannotti, and M. F. Kaashoek, "The Click Modular Router," *ACM Trans. Comput. Syst.*, vol. 18, no. 3, pp. 263–297, Aug. 2000.

[62] R. Tynan, G. M. P. O'Hare, M. J. O'Grady, and C. Muldoon, "Virtual Sensor Networks: An Embedded Agent Approach," in *International Symposium on Parallel and Distributed Processing with Applications, 2008. ISPA '08*, 2008, pp. 926–932.

[63] C. Muldoon, G. M. P. O'Hare, and J. F. Bradley, "Towards Reflective Mobile Agents for Resource-constrained Mobile Devices," in *Proceedings of the 6th International Joint Conference on Autonomous Agents and Multiagent Systems*, New York, NY, USA, 2007, pp. 141:1–141:3.

[64] H. M. N. D. Bandara, A. P. Jayasumana, and T. H. Illangasekare, "Cluster Tree Based Self Organization of Virtual Sensor Networks," in *2008 IEEE GLOBECOM Workshops*, 2008, pp. 1–6.

[65] Q. Han, A. P. Jayasumana, T. Illangaskare, and T. Sakaki, "A wireless sensor network based closed-loop system for subsurface contaminant plume monitoring," in *IEEE International Symposium on Parallel and Distributed Processing*, 2008. IPDPS 2008, 2008, pp. 1–5.

[66] M. Haghighi and D. Cliff, "Multi-agent Support for Multiple Concurrent Applications and Dynamic Data-Gathering in Wireless Sensor Networks," in *2013 Seventh International Conference on Innovative Mobile and Internet Services in Ubiquitous Computing (IMIS)*, 2013, pp. 320–325.

[67] R. B. Smith, "SPOTWorld and the Sun SPOT," in *Proceedings of the 6th International Conference on Information Processing in Sensor Networks*, New York, NY, USA, 2007, pp. 565–566.

[68] A. Majeed and T. A. Zia, "Multi-set Architecture for Multi-applications Running on Wireless Sensor Networks," in *2010 IEEE 24th International Conference on Advanced Information Networking and Applications Workshops (WAINA)*, 2010, pp. 299–304.

[69] J. Steffan, L. Fiege, M. Cilia, and A. Buchmann, "Towards multi-purpose wireless sensor networks," in *Systems Communications*, 2005. Proceedings, 2005, pp. 336–341.

[70] I. Leontiadis, C. Efstratiou, C. Mascolo, and J. Crowcroft, "SenShare: Transforming Sensor Networks into Multi-application Sensing Infrastructures," in *Wireless Sensor Networks*, G. P. Picco and W. Heinzelman, Eds. Springer Berlin Heidelberg, 2012, pp. 65–81.

[71] J. W. Hui and D. Culler, "The Dynamic Behavior of a Data Dissemination Protocol for Network Programming at Scale," in *Proceedings of the 2nd International Conference on Embedded Networked Sensor Systems*, New York, NY, USA, 2004, pp. 81–94.

[72] J. Lu, et al., "The Smart Thermostat: Using Occupancy Sensors to Save Energy in Homes," in *Proceedings of the 8th ACM Conference on Embedded Networked Sensor Systems*, New York, NY, USA, 2010, pp. 211–224.

[73] O. Gnawali, R. Fonseca, K. Jamieson, D. Moss, and P. Levis, "Collection Tree Protocol," in *Proceedings of the 7th ACM Conference on Embedded Networked Sensor Systems*, New York, NY, USA, 2009, pp. 1–14.

[74] M. Navarro, M. Antonucci, L. Sarakis, and T. Zahariadis, "VITRO Architecture: Bringing Virtualization to WSN World," in *2011 IEEE 8th International Conference on Mobile Adhoc and Sensor Systems (MASS)*, 2011, pp. 831–836.

[75] T. Zahariadis, P. Trakadas, H. C. Leligou, S. Maniatis, and P. Karkazis, "A Novel Trust-Aware Geographical Routing Scheme for Wireless Sensor Networks," *Wireless Pers Commun*, vol. 69, no. 2, pp. 805–826, Mar. 2013.

[76] P. Levis, "Experiences from a Decade of TinyOS Development," in *Proceedings of the 10th USENIX Conference on Operating Systems Design and Implementation*, Berkeley, CA, USA, 2012, pp. 207–220.

[77] R. N. Murty, G. Mainland, I. Rose, A. R. Chowdhury, A. Gosain, J. Bers, and M. Welsh, "CitySense: An Urban-Scale Wireless Sensor Network and Testbed," in *2008 IEEE Conference on Technologies for Homeland Security*, 2008, pp. 583–588.

[78] C. Efstratiou, I. Leontiadis, C. Mascolo, and J. Crowcroft, "A Shared Sensor Network Infrastructure," in *Proceedings of the 8th ACM Conference on Embedded Networked Sensor Systems*, New York, NY, USA, 2010, pp. 367–368.

[79] L. Sanchez, et al., "SmartSantander: IoT experimentation over a smart city testbed," *Computer Networks*, vol. 61, pp. 217–238, Mar. 2014.

[80] F. Berkers, et al., "Constructing a multi-sided business model for a smart horizontal IoT service platform," in *2013 17th International Conference on Intelligence in Next Generation Networks (ICIN)*, 2013, pp. 126–132.

[81] A. Andrushevich, B. Copigneaux, R. Kistler, A. Kurbatski, F. L. Gall, and A. Klapproth, "Leveraging Multi-domain Links via the Internet of Things," in *Internet of Things, Smart Spaces, and Next Generation Networking*, S. Balandin, S. Andreev, and Y. Koucheryavy, Eds. Springer Berlin Heidelberg, 2013, pp. 13–24.

[82] Irwin, David, et al. "Towards a virtualized sensing environment," in *Testbeds and Research Infrastructures. Development of Networks and Communities*. Springer Berlin Heidelberg, 2011, pp. 133-142.

[83] R. S. Oliver, I. Shcherbakov, and G. Fohler, "An Operating System Abstraction Layer for Portable Applications in Wireless Sensor Networks," in *Proceedings of the 2010 ACM Symposium on Applied Computing*, New York, NY, USA, 2010, pp. 742–748.

[84] J. Mäenpää, J. J. Bolonio, and S. Loreto, "Using RELOAD and CoAP for wide area sensor and actuator networking," *J Wireless Com Network*, vol. 2012, no. 1, pp. 1–22, Dec. 2012.

[85] Z. Shelby, "Embedded web services," *IEEE Wireless Communications*, vol. 17, no. 6, pp. 52–57, Dec. 2010.

[86] Shelby, Z., et al. "Constrained Application Protocol (CoAP), draft-ietf-core-coap-18", work in progress. *The Internet Engineering Task Force–IETF*, June (2013).

[87] Cheshire, S. and M. Krochmal, "Multicast DNS", *IETF RFC 6762*, February 2013.

[88] R. Glitho, M. Morrow, and P. Polakos, "A cloud based — Architecture for cost-efficient applications and services provisioning in wireless sensor networks," in *Wireless and Mobile Networking Conference (WMNC), 2013 6th Joint IFIP*, 2013, pp. 1–4.

[89] H. Rowaihy, M. P. Johnson, O. Liu, A. Bar-Noy, T. Brown, and T. L. Porta, "Sensor-mission Assignment in Wireless Sensor Networks," *ACM Trans. Sen. Netw.*, vol. 6, no. 4, pp. 36:1–36:33, Jul. 2010.

[90] M. Haghighi, "Market-Based Resource Allocation for Energy-Efficient Execution of Multiple Concurrent Applications in Wireless Sensor Networks," in *Mobile, Ubiquitous, and Intelligent Computing*, Jong H. Park, H. Adeli, N. Park, and I. Woungang, Eds. Springer Berlin Heidelberg, 2014, pp. 173–178.

[91] W. Li, F. C. Delicato, P. F. Pires, and A. Y. Zomaya, "Energy-efficient task allocation with quality of service provisioning for concurrent applications in multi-functional wireless sensor network systems," *Concurrency Computat.: Pract. Exper.*, vol. 26, no. 11, pp. 1869–1888, Aug. 2014.7

[92] J. Ko, B.-B. Lee, S. G. Hong, and N. Kim, "Poster Abstract: Virtualizing External Wireless Sensors for Designing Personalized Smartphone Services," in *Proceedings of the 12th International Conference on Information Processing in Sensor Networks*, New York, NY, USA, 2013, pp. 353–354.

[93] V. Gupta, N. Pereira, E. Tovar, and R. (Raj) Rajkumar, "Poster Abstract: A Harmony of Sensors: Achieving Determinism in Multi-application Sensor Networks," in *Proceedings of the 13th International Symposium on Information Processing in Sensor Networks*, Piscataway, NJ, USA, 2014, pp. 299–300.

[94] C.-M. Hsieh, Z. Wang, and J. Henkel, "DANCE: Distributed Application-aware Node Configuration Engine in Shared Reconfigurable Sensor Networks," in *Proceedings of the Conference on Design, Automation and Test in Europe*, San Jose, CA, USA, 2013, pp. 839–842.

[95] T. M. Cao, B. Bellata, and M. Oliver, "Design of a generic management system for wireless sensor networks," *Ad Hoc Networks*, vol. 20, pp. 16–35, Sep. 2014.

[96] Haghighi, Mo. "An Agent-based Multi-model Tool for Simulating Multiple Concurrent Applications in WSNs." Journal of Advances in Computer Networks (JACN), 5th International Conference on Communication Software and Networks, Malaysia (June 2013). 2013.

[97] C. Reed, et al. "Ogc® sensor web enablement:overview and high level achhitecture.," in *2007 IEEE Autotestcon*, 2007, pp. 372–380.

[98] E. Gelenbe, D. Gesbert, D. Gunduz, H. Kulah, and E. Uysal-Biyikoglu, "Energy harvesting communication networks: Optimization and demonstration (the E-CROPS project)," in *2013 24th Tyrrhenian International Workshop on Digital Communications - Green ICT (TIWDC)*, 2013, pp. 1–6.

[99] S. Sudevalayam and P. Kulkarni, "Energy Harvesting Sensor Nodes: Survey and Implications," *IEEE Communications Surveys Tutorials*, vol. 13, no. 3, pp. 443–461, Sep. 2011.

**IMRAN KHAN** (imran@ieee.org) received his BCS degree in Computer Science from COMSATS Institute of IT, Pakistan in 2005 and M.S. degree in Multimedia and Communication from M.A. Jinnah University, Pakistan in 2009. Currently he is a Ph.D. research student at Institut Mines-Télécom, Télécom SudParis jointly with Paris VI (UPMC). He is also a collaborating researcher at Concordia University, Montreal working on a CISCO project. In past he worked on projects funded by ISOC and ITEA2. He is student member of IEEE. His research interests are virtualization, cloud computing, wireless sensor networks, Internet of Thing (IoT), and M2M communications.

**FATNA BELQASMI** (fatna.belqasmi@zu.ac.ae) holds a Ph.D. and an M.Sc. degree in electrical and computer engineering from Concordia University, Canada. She is current working as Assistant Professor at Zayed University Abu Dhabi, UAE. In the past, she worked as a research associate at Concordia University, Canada and as a researcher at Ericsson Canada. She was part of the IST Ambient Network project (a research project sponsored by the European Commission within the Sixth Framework Programme -FP6-). She worked as an R&D engineer for Maroc Telecom in Morocco. Her research interests include next generation networks, service engineering, distributed systems, and networking technologies for emerging economies.

**ROCH GLITHO** [SM] (http://users.encs.concordia.ca/~glitho/) holds a Ph.D. (Tekn. Dr.) in tele-informatics (Royal Institute of Technology, Stockholm, Sweden) and M.Sc. degrees in business economics (University of Grenoble, France), pure mathematics (University Geneva, Switzerland), and computer science (University of Geneva). He works in Montreal, Canada, as an associate professor of networking and telecommunications at the Concordia Institute of Information Systems Engineering (CIISE) where he leads the telecommunication service engineering (TSE) research laboratory (.http://users.encs.concordia.ca/~tse/). In the past he has worked in industry for almost a quarter of a century and has held several senior technical positions at LM Ericsson in Sweden and Canada (e.g. expert, principal engineer, senior specialist). His industrial experience includes research, international standards setting (e.g. contributions to ITU-T, ETSI, TMF, ANSI, TIA, and 3GPP), product management, project management, systems engineering and software/firmware design. In the past he has served as IEEE Communications Society distinguished lecturer, Editor-In-Chief of IEEE Communications Magazine and Editor-In-Chief of IEEE Communications Surveys & Tutorials. His research areas are: virtualization and cloud computing; Machine-to-Machine communications (M2M) and Internet of Things; Distributed systems (e.g. SOAP Based – Web Services, RESTful Web Services); Rural communications and networking technologies for emerging economies.

**NOEL CRESPI** (noel.crespi@mines-telecom.fr) holds a Master's from the Universities of Orsay and Kent, a diplome d'ingénieur from Telecom ParisTech, and a Ph.D. and a Habilitation from Paris VI University. He worked from 1993 in CLIP, Bouygues Telecom, France Telecom R&D in 1995, and Nortel Networks in 1999. He joined Institut Mines-Télécom in 2002 and is currently professor and program director, leading the Service Architecture Laboratory. He is appointed as coordinator for the standardization activities in ETSI and 3GPP. He is also a visiting professor at the Asian Institute of Technology and is on the four-person Scientific Advisory Board of FTW, Austria. His current research interests are in service architectures, P2P service overlays, future Internet, and Web-NGN convergence. He is the author/co-author of more than 230 papers and contributions in standardization.

**MONIQUE MORROW** (mmorrow@cisco.com) holds the title of CTO Cisco Services. Ms. Morrow's focus is in developing strategic technology and business architectures for Cisco customers and partners. With over 13 years at Cisco, Monique has made significant contributions in a wide range of roles, from Customer Advocacy to Corporate Consulting Engineering. With particular emphasis on the Service Provider segment, her experience includes roles in the field (Asia-Pacific) where she undertook the goal of building a strong technology team, as well as identifying and grooming a successor to assure a smooth transition and continued excellence. Monique has consistently shown her talent for forward thinking and risk taking in exploring market opportunities for Cisco. She was an early visionary in the realm of MPLS as a technology service enabler, and she was one of the leaders in developing new business opportunities for Cisco in the Service Provider segment, SP NGN. Monique holds 3 patents, and has an additional nine patent submissions filed with US Patent Office. Ms. Morrow is the co-author of several books, and has authored numerous articles. She also maintains several technology blogs, and is a major contributor to Cisco's Technology Radar, having achieved Gold Medalist Hall of Fame status for her contributions. Monique is also very active in industry associations. She is a new member of the Strategic Advisory Board for the School of Computer Science at North Carolina State University. Monique is particularly passionate about Girls in ICT and has been active at the ITU on this topic - presenting at the EU Parliament in April of 2013 as an advocate for Cisco. Within the Office of the CTO, first as an individual contributor, and now as CTO, she has built a strong leadership team, and she continues to drive Cisco's globalization and country strategies.

**PAUL POLAKOS** (ppolakos@cisco.com) is currently a Cisco Fellow and member of the Mobility CTO team at Cisco Systems focusing on emerging technologies for future Mobility systems. Prior to joining Cisco, Paul was Senior Director of Wireless Networking Research at Bell Labs, Alcatel-Lucent in Murray Hill, NJ and Paris, France. During his 28 years at Bell Labs he worked on a broad variety of topics in Physics and in Wireless Networking Research including the flat-IP cellular network architecture, the Base Station Router, femtocells, intelligent antennas and MIMO, radio and modem algorithms and ASICSs, autonomic networks and dynamic network optimization. Prior to joining Bell Labs, he was a member of the research staff at the Max-Planck Institute for Physics and Astrophysics (Munich) and visiting scientist at CERN and Fermilab. He holds BS, MS, and Ph.D. degrees in Physics from Rensselaer Polytechnic Institute and the University of Arizona, is a Bell Labs and Cisco Fellow, and author of more than 50 publications and 30 patents.